\documentclass[intlimits,twoside,a4paper]{article}

\usepackage[cp1251]{inputenc}

\usepackage{graphicx}

\usepackage{cmpj3}
\usepackage{bm}

\issue{2020}{23}{2}{23602}
\doinumber{10.5488/CMP.23.23602}

\title[Methanolic solution of NaCl]{On the properties of methanolic NaCl solution by 
molecular dynamics simulations\thanks{This article is dedicated to Prof. I. Mryglod on the occasion of
his 60th birthday.}}

\author[M. Cruz Sanchez, H. Dominguez, O. Pizio]{M. Cruz Sanchez\refaddr{label1}, H. Dominguez\refaddr{label2}, O. Pizio\refaddr{label3}\footnote{Corresponding author, oapizio@gmail.com.}}

\addresses{
\addr{label1} Instituto de Qu\'{i}mica, Universidad Nacional Aut\'{o}noma de M\'{e}xico,
Circuito Exterior, 04510, Cd. de M\'{e}xico, M\'{e}xico
\addr{label2} Instituto de Investigaciones en Materiales, Universidad Nacional 
Aut\'{o}noma de M\'{e}xico, Circuito Exterior, 04510, Cd. de M\'{e}xico, M\'{e}xico
\addr{label3} Instituto de Qu\'{i}mica, Universidad Nacional Aut\'{o}noma de M\'{e}xico,
Circuito Exterior, 04510, Cd. de M\'{e}xico, M\'{e}xico
}

\date{Received January 30, 2020, in final form February 21, 2020}

\begin{document}
\maketitle

\begin{abstract}
Isothermal-isobaric  molecular dynamics simulations 
are used to examine the microscopic structure and principal thermodynamic properties
of a model solution consisting of NaCl salt dissolved in methanol solvent. 
Four united atom force fields for methanol are involved. 
Concerning ion solutes we used the
Joung-Cheatham, Smith-Dang models as well as the model from the laboratory of Vrabec.
Our principal focus is to evaluate the quality of predictions of
different combinations of models for basic properties of these solutions.
Specifically, we explored the change of density on molality, the 
structural properties in terms of various pair distribution functions, 
the coordination numbers, the number of ion pairs and the average number of hydrogen bonds.
In addition,  changes of the self-diffusion coefficients of species, the solvent
dielectric constant and the evolution 
of the surface tension  with ion concentration are described.


\end{abstract}

\section{Introduction}

From a general perspective, this work is focused on the theoretical description of one simplest example
of a class of electrolyte systems that comprise ionic solutes dissolved in water with 
organic co-solvent. Molecular dynamics computer simulations are used as tools to 
explore the microscopic structure, ions solvation and thermodynamic properties of interest.

This kind of systems exhibit a rich variety of physicochemical phenomena and are
of much interest for basic research. Besides, the subject is of much importance for several 
electrochemical applications, see e.g.,~\cite{kang-xu}. 
Various ionic solutions involving different co-solvents have been investigated using
experimental techniques and computer simulations~\cite{patey4,tembe,ewa1,ewa2,ewa3,ewa4,ewa5,ewa6,ewa7,ewa8,ewa9,takamuku1,takamuku2,takamuku3,takamuku4,takamuku5,taha,bouazizi,chapman}.

The present contribution is a part of our ongoing project focused on the theoretical exploration
of properties of ionic solutes in water-alcohol solvents of variable composition.
At the first stage, water-methanol model
mixtures have been explored in this laboratory~\cite{galicia1,galicia2,galicia3,cruz2}, 
continuing several previous studies of this kind of systems, see 
e.g.,~\cite{ferrario,gabor,wensink,ivan-ortega,laaksonen,guevara,bopp2,lomba} and respective references
in the above cited works. Still, there remains room for additional insights, 
having in mind the recent novel modelling of liquid methanol~\cite{vega,cuernavaca}.

One limiting case of ionic solutes in water without organic co-solvent generated
huge amount of computer simulation results during past decades. 
Several remaining problems within this methodology are due to a limited understanding 
of water as solvent on its own, and of the influence of
ions on water behaviour. For example, modelling of water in the framework of a
particular  model makes necessary to use adequate, specific set of parameters describing 
force field for ionic solutes. This issue is well documented for example in the development 
of frequently used Joung-Cheatham (JC)~\cite{jcc} model for alkali halides aqueous solutions
with TIP3P~\cite{tip3p}, TIP4P-Ew~\cite{tip4p-ew} and SPC/E~\cite{spce} water models.
Methodological aspects of the development and application of ionic force fields
to NaCl aqueous electrolytes have been critically reviewed in 
\cite{ivo-1,ivo-2,ivo-3,vrabec1}. The state of art for this particular 
aqueous solution is well described in~\cite{benavides1,benavides2,alejandre}.

On the other hand, the most comprehensive investigation of ionic solutions in water-methanol 
solvents was performed in the laboratory of 
 Hawlicka~\cite{ewa1,ewa2,ewa3,ewa4,ewa5,ewa6,ewa7,ewa8,ewa9}. These studies were
restricted to NaCl and NaI univalent salts, as well as MgCl$_2$ and CaCl$_2$ salts 
with divalent cations as solutes.
The solvent species was described by using the BJH water~\cite{bopp} and
PHH~\cite{palinkas} methanol models. In addition, the \textit{ab~initio} calculations for
ion-solvent interaction energies were involved.
Some recent contributions contributed to a better knowledge of the properties
of NaCl solutions in water-methanol solvent as well~\cite{kohns,clark}.

In contrast to aqueous NaCl solution, the non-aqueous  NaCl solution with pure methanol 
as solvent is much less studied. Even the amount of experimental data concerning the properties
of this system  is much less comprehensive.
A set of models for liquid methanol, designed for application within the computer
simulation methods, is quite 
ample~\cite{vega,cuernavaca,palinkas,haughney,jorgensen,leeuwen,vrabec2,trappe}.
As concerns computer simulation  studies of solutions of simple salts in methanol, we
are aware of the following reports~\cite{marx,karl,reiser,sese,chaban}.

Recently, in \cite{cruz1}, we explored the properties of NaCl solutions with
water-methanol mixed solvent using the SPC/E water~\cite{spce}, united atom
methanol~\cite{jorgensen} and NaCl force field due to Dang~\cite{dang}.
However, the limit of methanolic NaCl solution has not been explored profoundly.
Therefore, in the present work we would like to study this particular case more in  
detail. In addition, our intention is to provide insights into the dependence
of the results on the choice of models for methanol and for ionic solutes.
This issue has not been investigated so far, in contrast to pure 
methanol~\cite{vega,cuernavaca}.  In essence, this is the primary objective
of the present contribution. The isobaric-isothermal molecular dynamics computer 
simulations represent our tools.

\section{Models and simulation details}

In the present work,  we apply four united atom type models for methanol described in table~\ref{tab1}. 
The CH$_3$ group is treated as a single site denominated as C. The oxygen and hydroxyl hydrogen
sites are denoted as O and H, respectively.
More specifically, the methanol model from \cite{leeuwen},
the model developed in the laboratory of Vrabec~\cite{vrabec2}, the recently developed model from the 
laboratory of Vega~\cite{vega}, and the model from the TraPPE database~\cite{trappe} are used. They are
denominated as L1, L2, OPLS/2016, according to table~1 of \cite{vega}, 
and as TraPPE, respectively. The bond lengths and the bond angle for each model are given in table~\ref{tab1}.
  \begin{table}[!b]
    \caption{Parameters of Lennard-Jones interactions for the L1~\cite{leeuwen},
     L2~\cite{vrabec2}, OPLS/2016~\cite{vega}, and TraPPE~\cite{trappe} models for methanol. 
    All $\sigma$ are in \AA, $\varepsilon$ --- in kJ/mol  and charges, $q$, in $e$ units. 
    The L1 and L2 models assume the C--O--H angle at 108.53$^{\circ}$, whereas for the  
    OPLS/2016 and TraPPE models it is equal to 108.50$^{\circ}$. }
   \label{tab1}
    \vspace{2ex}
     \begin{center}
     \footnotesize
      \begin{tabular}{l c c c c c c c c c }
      \hline
       Model  &    $\varepsilon_\text{OO}$ & $\sigma_\text{OO}$ &    $\varepsilon_\text{CC}$ & $\sigma_\text{CC}$ &
                  $q_\text{O}$ & $q_\text{C}$ & $q_\text{H}$ & $d_\text{OH}$ & $d_\text{CO}$  \\
       \hline
       L1     &  0.719200 & 3.0300 & 0.874680 &  3.7400  & $-$0.70000 & 0.26500 & 0.43500 & 0.9451 & 1.4246  \\
       L2     &  0.730667 & 3.0300 & 1.002658 &  3.7543  & $-$0.67874 & 0.24746 & 0.43128 & 0.9451 & 1.4246  \\
    OPLS/2016 &  0.812947 & 3.1659 & 0.918333 &  3.6499  & $-$0.65440 & 0.15460 & 0.49980 & 0.9450 & 1.4300  \\
     TraPPE   &  0.773245 & 3.0200 & 0.814817 &  3.7500  & $-$0.70000 & 0.26500 & 0.43500 & 0.9450 & 1.4300  \\
       \hline
  \end{tabular}
  \end{center}
\end{table}

In table~\ref{tab2} we present the force fields for NaCl employed in the present work.
Namely, we use the Joung-Cheatham (JC)~\cite{jcc} and  Smith and Dang (SD)~\cite{smith-dang}
models, in close similarity to the description of NaCl aqueous solutions~\cite{benavides1}.
The values of parameters for ions are taken from JC/SPC/E and SD/SPC/E design, see
supplementary material for \cite{benavides1}, whereas the crossed, ion-methanol, interaction
parameters are obtained using the Lorentz-Berthelot (LB) combination rules.
In essence, the three-site SPC/E water model~\cite{spce} in our procedure is substituted 
by the methanol models with three sites.
We are aware that the OPLS-2016 force field is designed to operate with the 
geometric combination  rules (CR3 according to the GROMACS nomenclature). In this work, however,
we applied solely the LB rules (CR2) in order to have a single and common benchmark for all the
models. 

In addition, we complement the L2 methanol model~\cite{vrabec2} with the model for NaCl
originally proposed in the work from the same laboratory~\cite{deublein} and later adjusted and applied to
describe the properties of aqueous and methanolic NaCl solutions in~\cite{reiser,vrabec3}. 
Again, the LB combination rules are used for this model.
In all these models, the ion valency is assumed equal to unity. 

\begin{table}[!t]
   \caption{Parameters of the Lennard-Jones interaction between ions for the JC~\cite{jcc},
  SD~\cite{smith-dang}, and VR~(Vrabec et al.)~\cite{reiser}  models.
  The diameters of ions, $\sigma$, are in {\AA}, and inter-particle interaction
  energies, $\varepsilon$, are in kJ/mol.}
    \label{tab2}
    \vspace{2ex}
\begin{center}
  \begin{tabular}{l| c  c| c c| c c }
      \hline
                      & \multicolumn{2}{c|}{JC}  & 
                        \multicolumn{2}{c|}{SD}  &    \multicolumn{2}{c}{VR} \\
            site      & $\sigma$ & $\varepsilon$  
                      & $\sigma$ & $\varepsilon$      & $\sigma$ & $\varepsilon$   \\
       \hline
           Cl$^-$     & 4.830  &  0.0535   &  4.400  &  0.4184 &  4.410  &  1.66207  \\
           Na$^+$     & 2.160  &  1.4754   &  2.350  &  0.5439 &  1.890  &  1.66207 \\
           \hline
  \end{tabular}
\end{center}
\end{table}

Our calculations were performed in the isothermal-isobaric ($NPT$) ensemble
at 1~bar, and at a temperature of 298.15~K. We used the GROMACS
software package~\cite{gromacs}, version 5.1.2.

As concerns the treatment of interactions for each model in question, we would like to mention
the following issues. Namely, the non-bonded interactions were cut off at 1.4~nm. 
The long-range electrostatic interactions were handled using the
particle mesh Ewald method implemented in the GROMACS software package (fourth
order, Fourier spacing equal to 0.12) with the precision $10^{-5}$.
The van der Waals tail correction terms to the energy and pressure were taken into account.
In order to maintain the geometry of the methanol  molecules, the LINCS algorithm was used.

On the other hand, concerning the procedure, a periodic cubic 
simulation box was set up for each system. 
The GROMACS genbox tool was employed to randomly place all particles in the 
simulation box. To remove possible overlaps of particles introduced by the procedure of
preparation
of the initial configuration, each system underwent energy
minimization using the steepest descent algorithm implemented in the GROMACS
package. Minimization was followed by a 50~ps $NPT$ equilibration run at 298.15~K and 1~bar using 
the timestep  0.25~fs.
We applied the Berendsen thermostat and
barostat with $\tau_T = 1$~ps and $\tau_P = 1$~ps during equilibration.
Constant value of 1.2$\times 10^{-4}$~bar$^{-1}$ for the compressibility of the solutions was
employed. 
The V-rescale thermostat and Parrinello-Rahman
barostat with $\tau_T = 0.5$~ps and $\tau_P = 2.0$~ps
and the time step 2~fs were used during production runs.

Statistics for various properties and various ion
concentrations was collected from a set of
several $NPT$ runs with duration of 10--20~ns, each started from the last configuration of the
preceding run. The time extension for each series of calculations will be mentioned below
in appropriate places, if necessary. The number of methanol solvent molecules
was fixed at 3000, the number of NaCl molecules (pairs of ions, denoted as $N_\text{ip}$) 
 was in the interval from 0 to 30.

\section{Results and discussion}

\subsection{Density of solutions}

\begin{figure}[!t]
\centering
\includegraphics[width=6.5cm,clip]{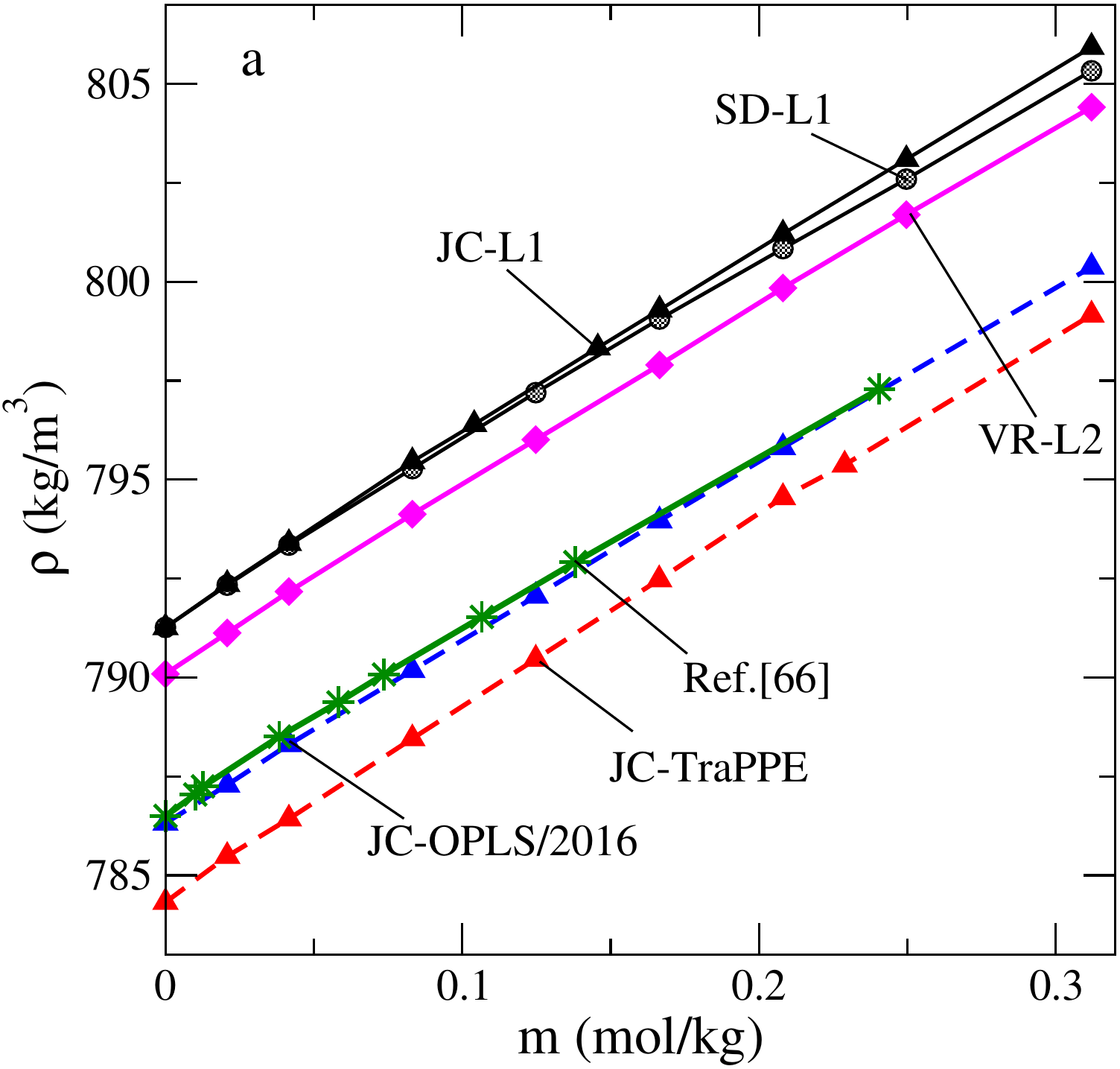}
\includegraphics[width=6.5cm,clip]{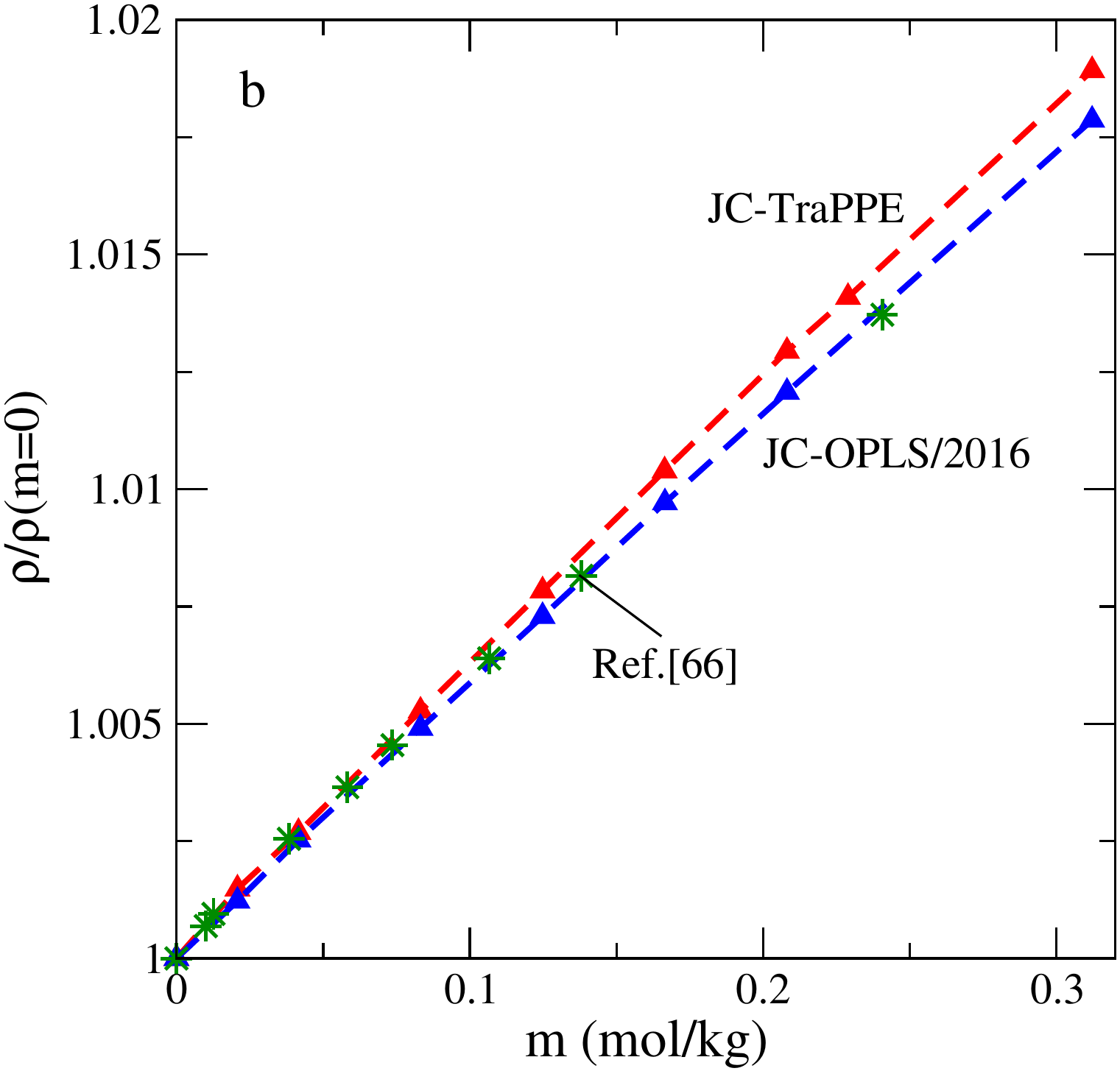}
\caption{(Colour online) 
Panel (a): Density of NaCl methanolic solutions on molality from $NPT$
simulations for SD-L1, JC-L1, VR-L2, JC-TraPPE, JC-OPLS/2016  models.
Experimental data are from~\cite{lankford}. 
Panel (b): Reduced density of the same models as in panel (a) on molality.
}
\label{fig1}
\end{figure}

Our first series of simulations are focused in the description of the dependence of density of
NaCl methanolic solution on solute molality, $m$,  in mol/kg units.
The simulation  results for the entire  set or models of this study are compared with experimental 
data from \cite{lankford}, in figure~\ref{fig1}~(a). Slightly less comprehensive, experimental results concerning 
the density of the system in question at room temperature can be found also in 
\cite{reiser,eliseeva,pasztor,takenaka}. The experimental solubility limit has been
established at $m=0.238$,~\cite{gmehling,pinho}.

Inspection of the results shown in figure~\ref{fig1}~(a) leads to the following conclusions. 
An overall location of the lines resulting from different models is determined by the
quality of description of the density of pure methanol.
All the models correctly describe trends of augmenting solution density on molality.
The growth seems to be almost linear.
The SD-L1, JC-L1, VR-L2 models overestimate the density of solution at different molalities 
whereas the JC-TraPPE model underestimates the density. The best agreement with the
experimental data is provided by the JC-OPLS/2016 model.
A comparison of simulation results for JC-L1 results with SD-L1 data illustrates the role of ions
force field. The effect of changing ions force field is small. Evidently, at low concentration 
of salt, the interactions between ions play minor role.
Their effect is slightly more pronounced at a high molality. 

Additional insight into the quality of presented results can be obtained from figure~\ref{fig1}~(b), 
in terms of reduced density on molality, i.e. the density of solution normalized by
the methanol density (this type of plot has been used in \cite{reiser}, cf. figure~5
of that work). 
In figure~\ref{fig1}~(b), we show the results for two models only that exhibit largest difference,
the curves from other models flow together, or say lay between the two curves shown in the figure. 
All the models in question perform well, but the JC-OPLS/2016 predictions seem to be the best.
Note that the scale of figure~\ref{fig1}~(b) is finer and more adequate, in comparison to the panel 
describing NaCl solution in figure~5 of ~\cite{reiser}.
The points of the curves from simulations at a highest molality  apparently should correspond 
to the over-saturated solutions according to the experiment data
(theoretical solubility of this salt in methanol has not been established so
far up to our best knowledge). However, analysis of the particles configurations 
by using the vmd software at these conditions ($N_\text{ip} = 30$) does not show 
peculiar features just demonstrating uniformly distributed ions throughout the 
simulation box.

Therefore, it is of interest to complement these insights by exploring 
changes of the microscopic structure of 
NaCl methanolic solutions upon increasing molality for different models in question.

\subsection{Pair distribution functions, coordination numbers and hydrogen bonding}

\begin{figure}[!t]
\centering
\includegraphics[width=6.55cm,clip]{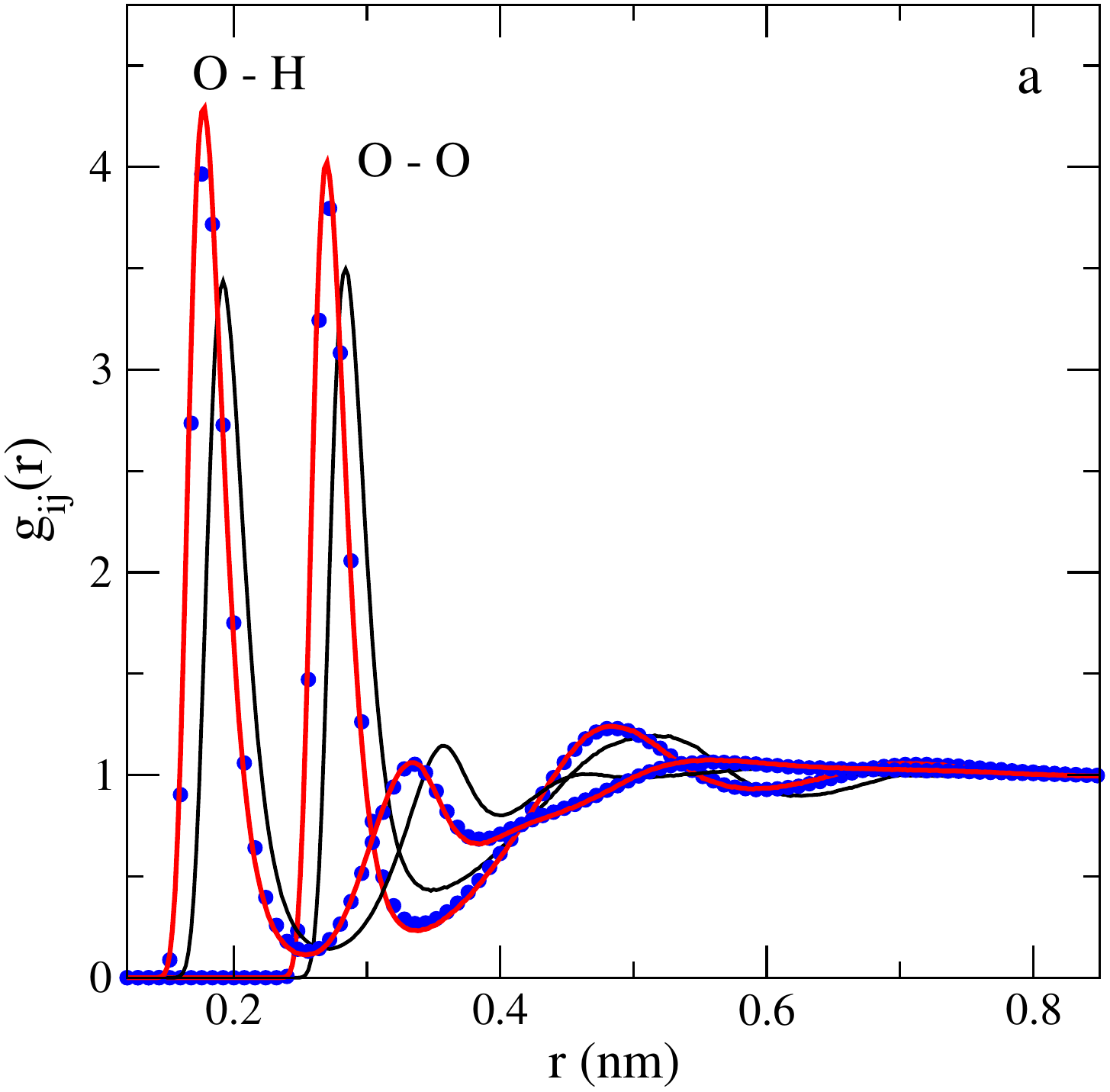}
\includegraphics[width=6.55cm,clip]{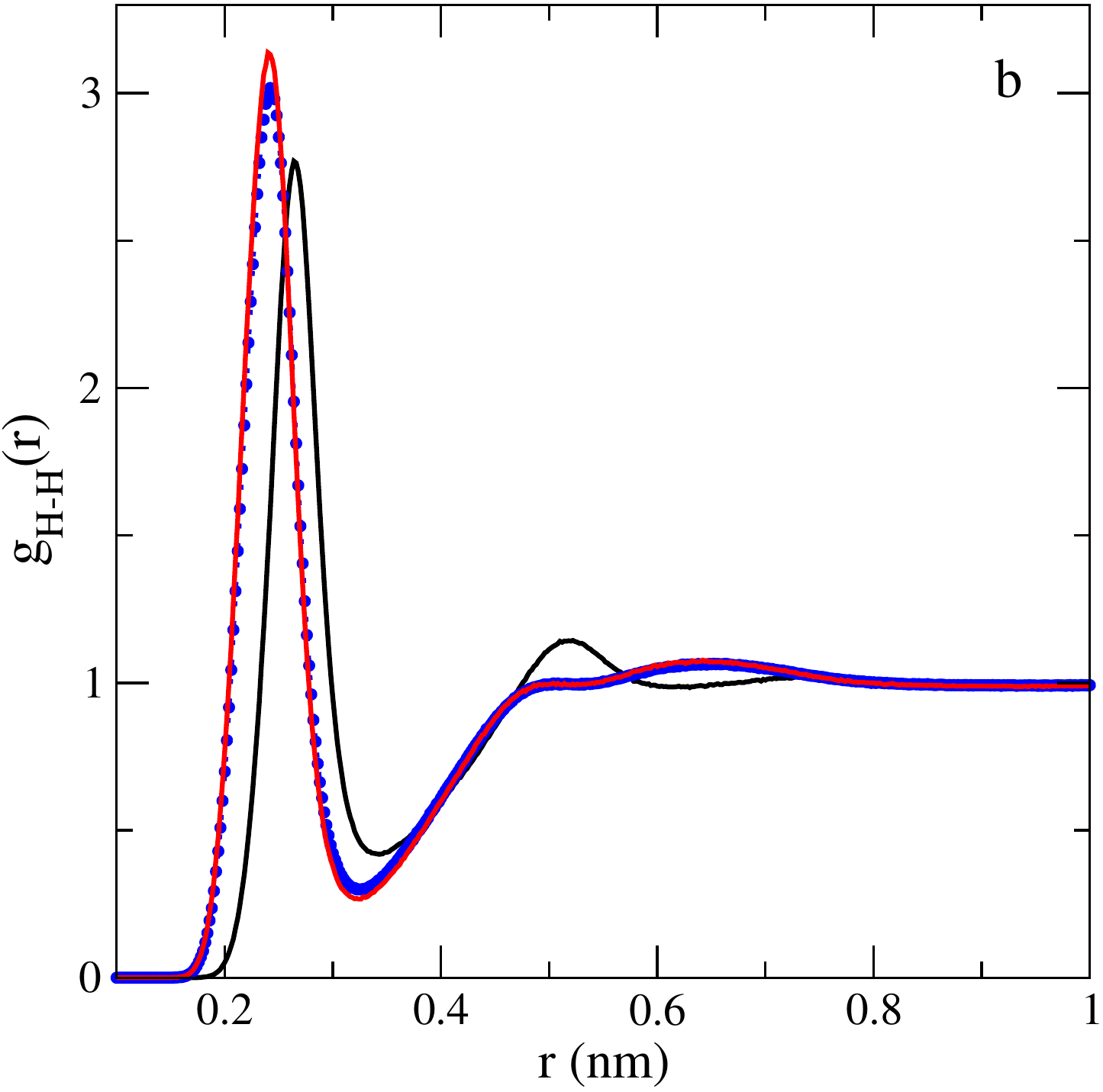}
\includegraphics[width=6.55cm,clip]{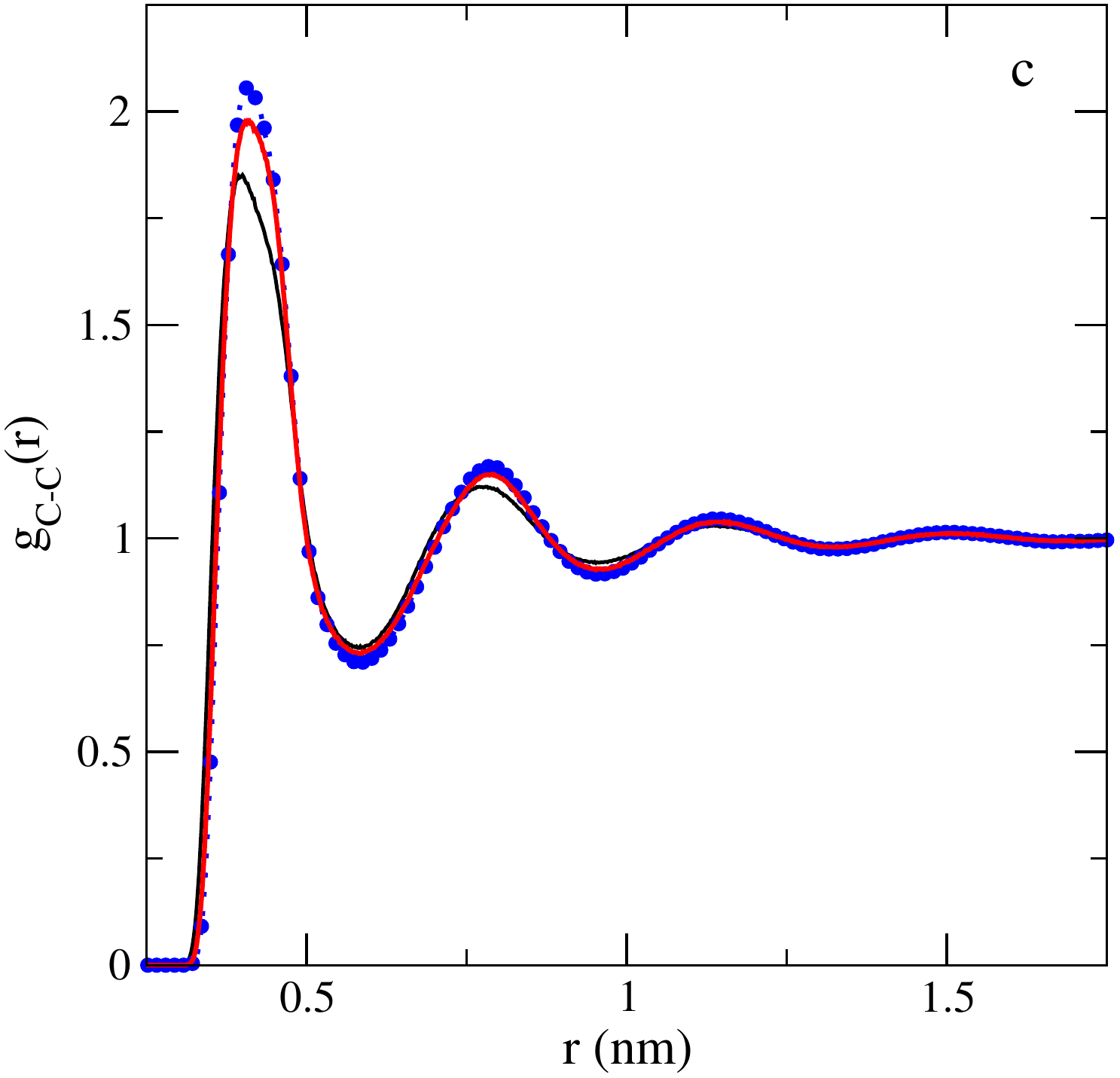}
\includegraphics[width=6.55cm,clip]{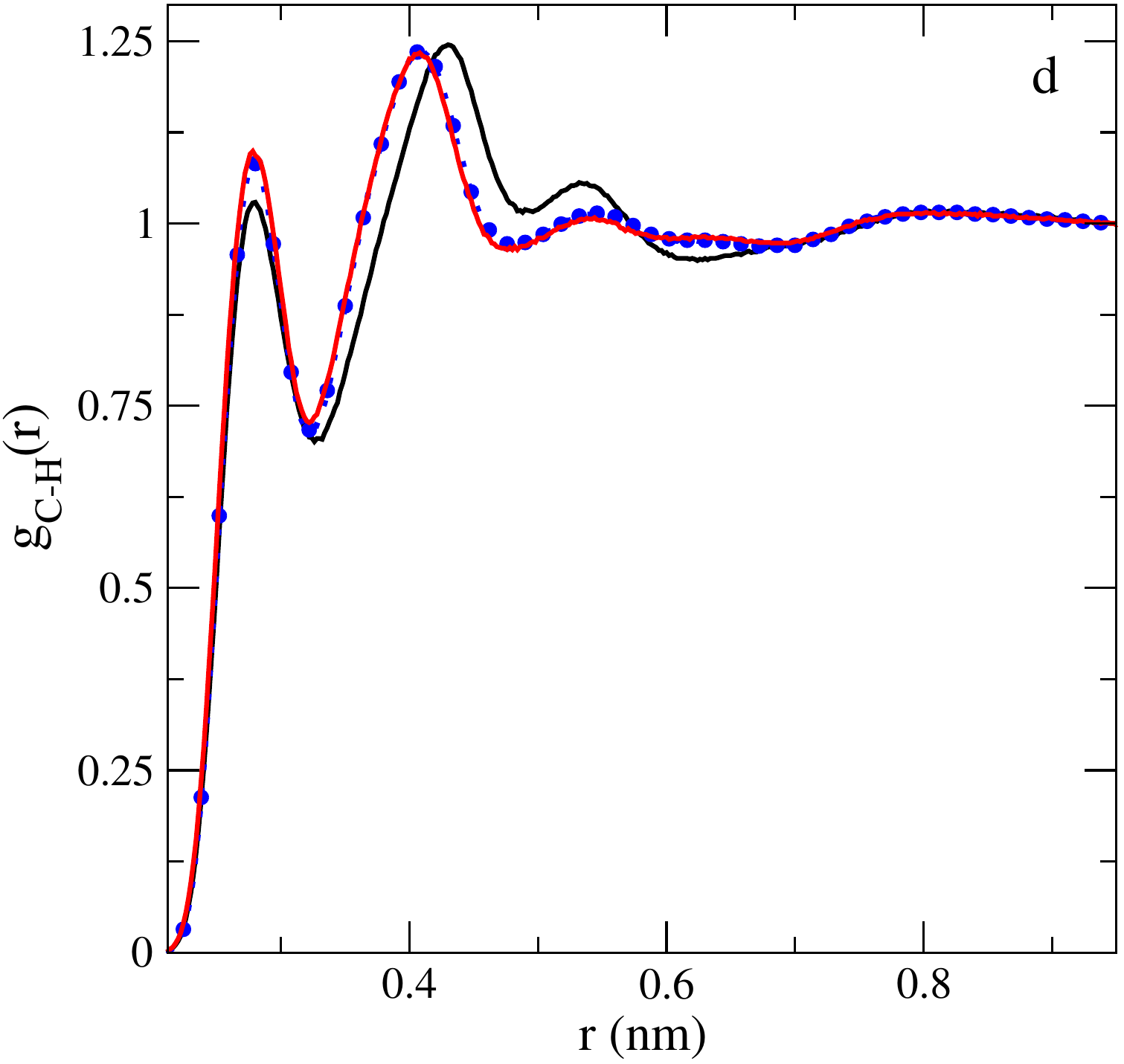}
\caption{(Colour online) 
Microscopic structure of pure methanol following from L2, TraPPE and OPLS/2016  models
in terms of the pair distribution functions, black solid line --- OPLS/2016,
red solid line --- TraPPE, blue circles --- L2 model, respectively.
}
\label{fig2}
\end{figure}

Evolution of the microscopic structure in the present work is described in terms of various
pair distribution functions (PDFs). The pair distribution functions of methanol species
have been discussed for various occasions. In order to avoid
unnecessary repetition, we would like to comment solely a few specific issues.
A set of PDFs describing the structure of pure methanol for three models,
the OPLS/2016, TraPPE and L2, is given in figure~\ref{fig2}. The curves for each kind of functions
resulting from different models are similar, the differences being observed
principally in the values of the first maxima, location of the first minima and the
phase of the following decaying oscillations. However, these, apparently, subtle differences
are manifested in more pronounced discrepancies for various thermodynamic 
and dynamic properties. 

In order to evaluate the quality of predictions of the 
microscopic structure we resort to the simulated PDFs for the L2 model in \cite{vrabec2}
(see figure~10). The PDFs from~\cite{vrabec2} have been compared with their counterparts 
extracted from the  neutron diffraction experiments and empirical potential structure 
refinement procedure by Yamaguchi et al.~\cite{yamaguchi1,yamaguchi2}. 
The quality of the agreement appeared to be very good.  Therefore, the PDFs for L2 model
can be considered as a useful reference.

The results from our simulations of the L2 model are given by blue lines and symbols 
in four panels of figure~\ref{fig2}.
The curves coming from the TraPPE model are very close to the results for L2 model.
Small differences can be observed only concerning the height of the first maximum for
O--O, O--H, H--H and C--C distributions. On the other hand, the functions
coming from the OPLS/2016 model essentially differ in predictions of the values of
the first maxima and minima and the phase of further oscillation compared to 
L2 and TraPPE. It is worth mentioning, however, that the OPLS/2016 predictions are close
to the results coming from PHH methanol model~\cite{palinkas}, as it follows
from the inspection of figure~5 shown in~\cite{bopp2}.

\begin{figure}[!t]
\centering
\includegraphics[width=6.5cm,clip]{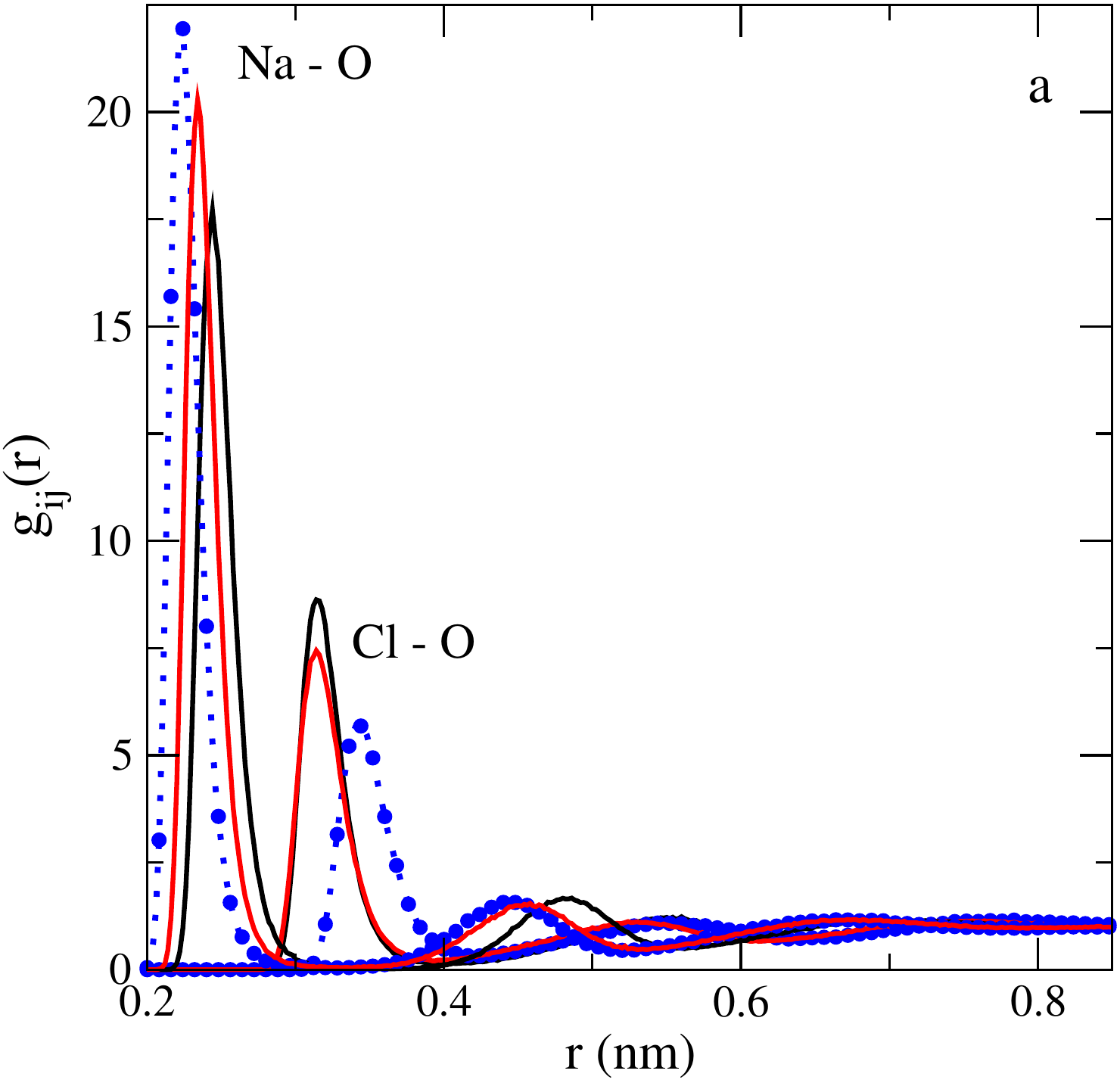}
\includegraphics[width=6.5cm,clip]{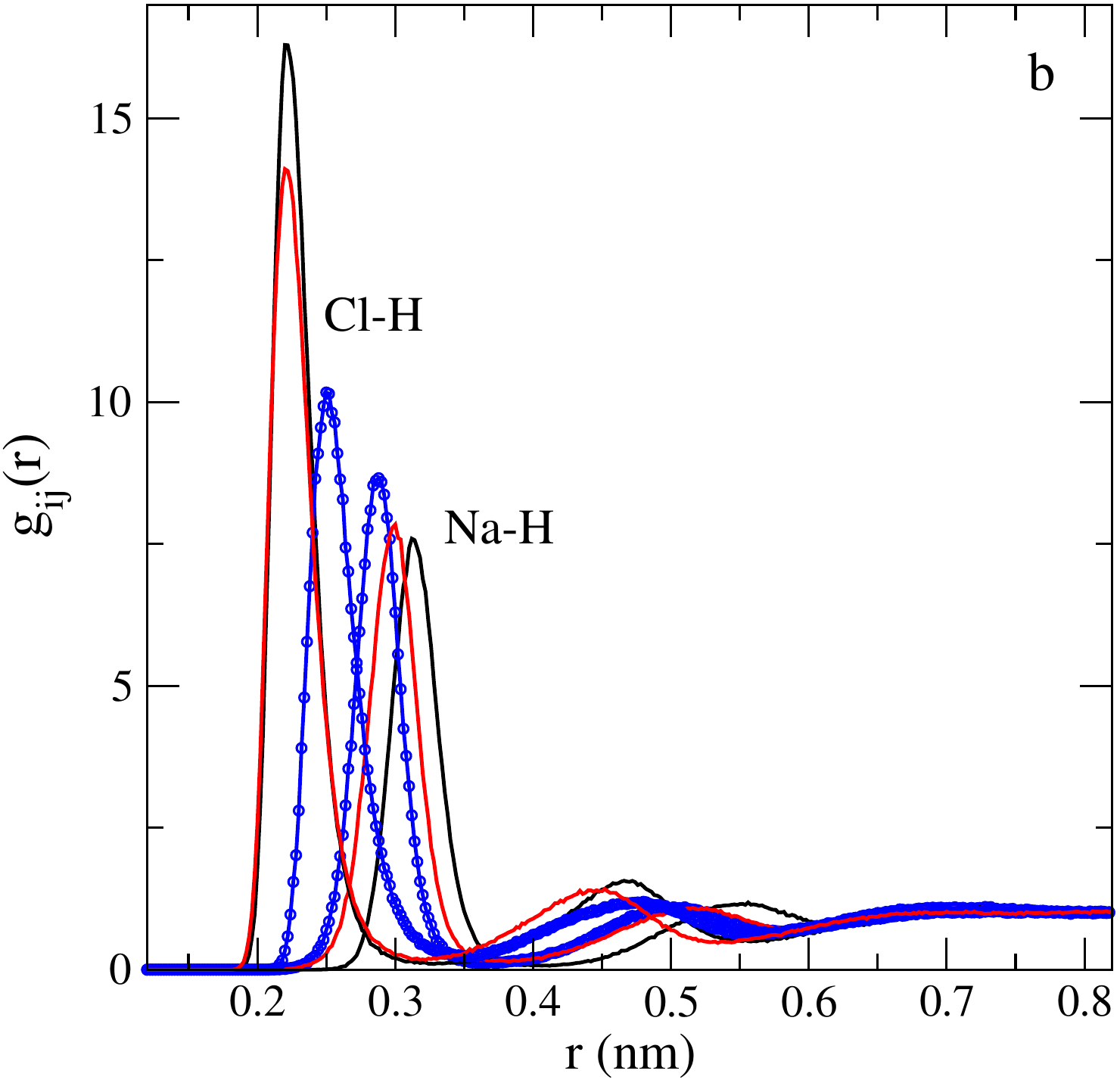}
\includegraphics[width=6.5cm,clip]{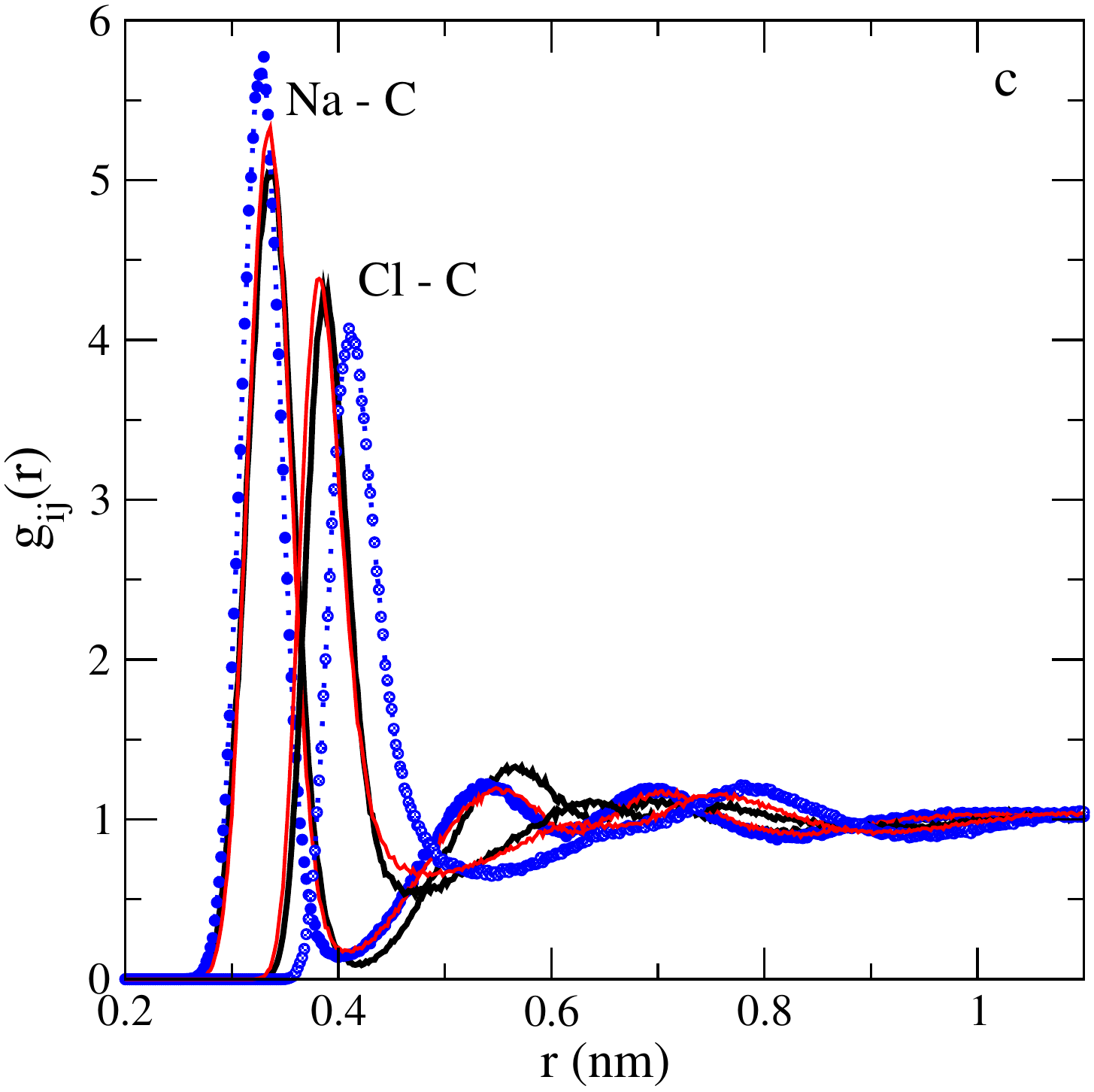}
\caption{(Colour online)  
Microscopic structure of NaCl methanol solutions following from the VR-L2, JC-TraPPE 
and JC-OPLS/2016 models in terms of ion-solvent pair distribution functions
at molality $m=0.0208$ (black solid line --- OPLS/2016, red solid line --- TraPPE, blue 
dotted line with circles --- L2 model, respectively). 
}
\label{fig3}
\end{figure}

The PDFs describing the distribution of solvent species around cation and anion,
from different models of the present study, is given in figure~\ref{fig3} for the lowest
molality studied, $m=0.0208$. For these functions, the differences between 
the predictions from different models are more pronounced compared to the previous figure. 
The sequence of positions and the heights of the first maximum of the function
Na--O is as follows: VR-L2 (the closest and highest), JC-TraPPE and
JC-OPLS/2016 (most distant and lowest). 
The Na$^+$ cation possesses a well pronounced 
first and second coordination shell. In order to describe mutual coordination of
the species, we use the notion of the
running coordination number as common,
\begin{equation}
 n_{i}(R)=4\piup\rho_j\int_{0}^{R}g_{ij}(r)r^{2}\rd r,
\end{equation}
where $\rho_j$ is the number density of species $j$. The first coordination number,
for example, is obtained  by putting $R=r_\text{min}$, i.e., at the first minimum of the
corresponding pair distribution function.

As it follows from the $g_\text{Na--O}(r)$ PDF, the first coordination number for Na$^+$ is: 5.5 (VR-L2),
5.7 (JC-TraPPE) and 5.9 (JC-OPLS/2016), respectively. The number of 
methanol solvent molecules in the second shell is obtained by resting
the first coordination number from the second coordination number. It is as follows: 
5.4 (VR-L2), 5.6 (JC-TraPPE) and 6.5 (JC-OPLS/2016), respectively. 
Our result for the VR-L2 model 
for the first coordination number compares well with the value 5.3 reported
for Na$^+$ recently at molality $m \approx 0.0313$ (cf. table 9 of \cite{reiser}). 
On the other hand, the solvation numbers 5.8 and 6.2, for the first coordination number and 
the number of methanols in the second coordination shell, have been reported in~\cite{marx} 
for the \textit{ab initio} type NaCl solute in PHH methanol for 0.6 molal solution.
Moreover, the first coordination number reported here for three models is in
agreement with the estimate of 5.7 deduced from IR overtone spectroscopy data
in \cite{robinson} and discussed previously in~\cite{marx}. 

The solvation of Cl$^-$ anion in methanol can be interpreted by using the
Cl--O, Cl--H and Cl--C distribution functions from figure~\ref{fig3}. 
Only the first coordination shell of Cl$^-$ is well pronounced, in contrast to Na$^+$.
Again, the JC-TraPPE and JC-OPLS/2016 models behave quite similar, just yielding slightly different
heights of the first maxima and a different phase of decaying oscillations. The VR-L2
model apparently predicts a more distant position of methanol molecules around the
anion. In quantitative terms, these features lead to the following first coordination
numbers from the $g_\text{Cl--O}(r)$ PDF: 5.5 (JC-TraPPE), 5.9 (JC-OPLS/2016) and 5.7 (VR-L2). These estimates are in
agreement with VR-L2 data yielding 5.5 (cf. table 9 of \cite{reiser}).
By contrast, the Cl$^-$ solvation with PHH methanol model yields a higher coordination
number of 7.2, \cite{marx}, compared to the models of the present study and
to the spectroscopy estimates \textit{ca.} 4, from \cite{robinson}. 
From the analysis of ion-methanol atoms distances according to the positions of
the first maxima of various PDFs in conjunction, one can construct a schematic picture
of cation and anion solvation by methanol. Two examples of this kind of plots
are given in figure~4 of~\cite{marx} and in figure~1 of~\cite{sese}. The models
described here in figure~\ref{fig3} provide a similar picture differing only in subtle details.

\begin{figure}[!b]
\centering
\includegraphics[width=5.9cm,clip]{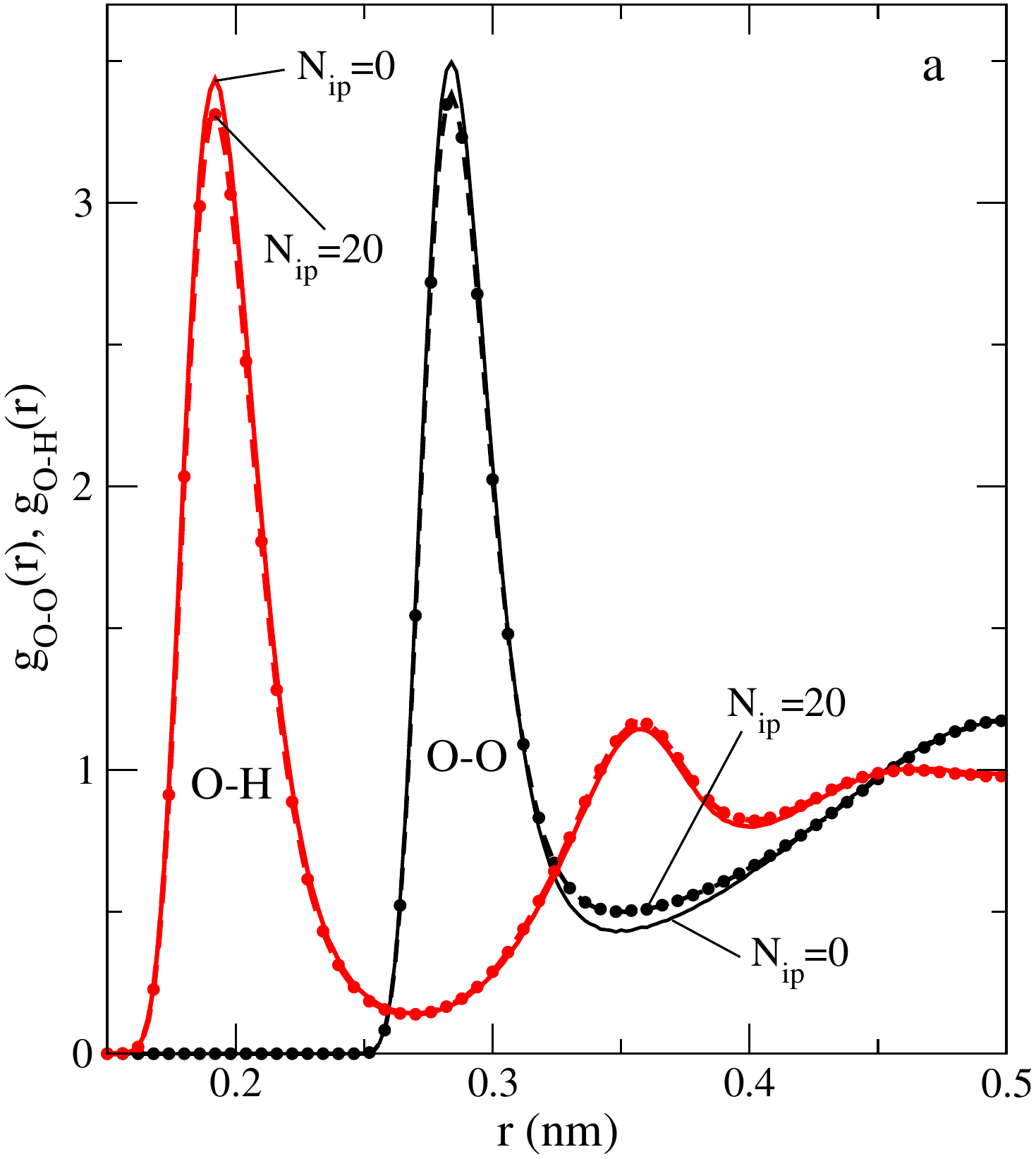}
\includegraphics[width=6.9cm,clip]{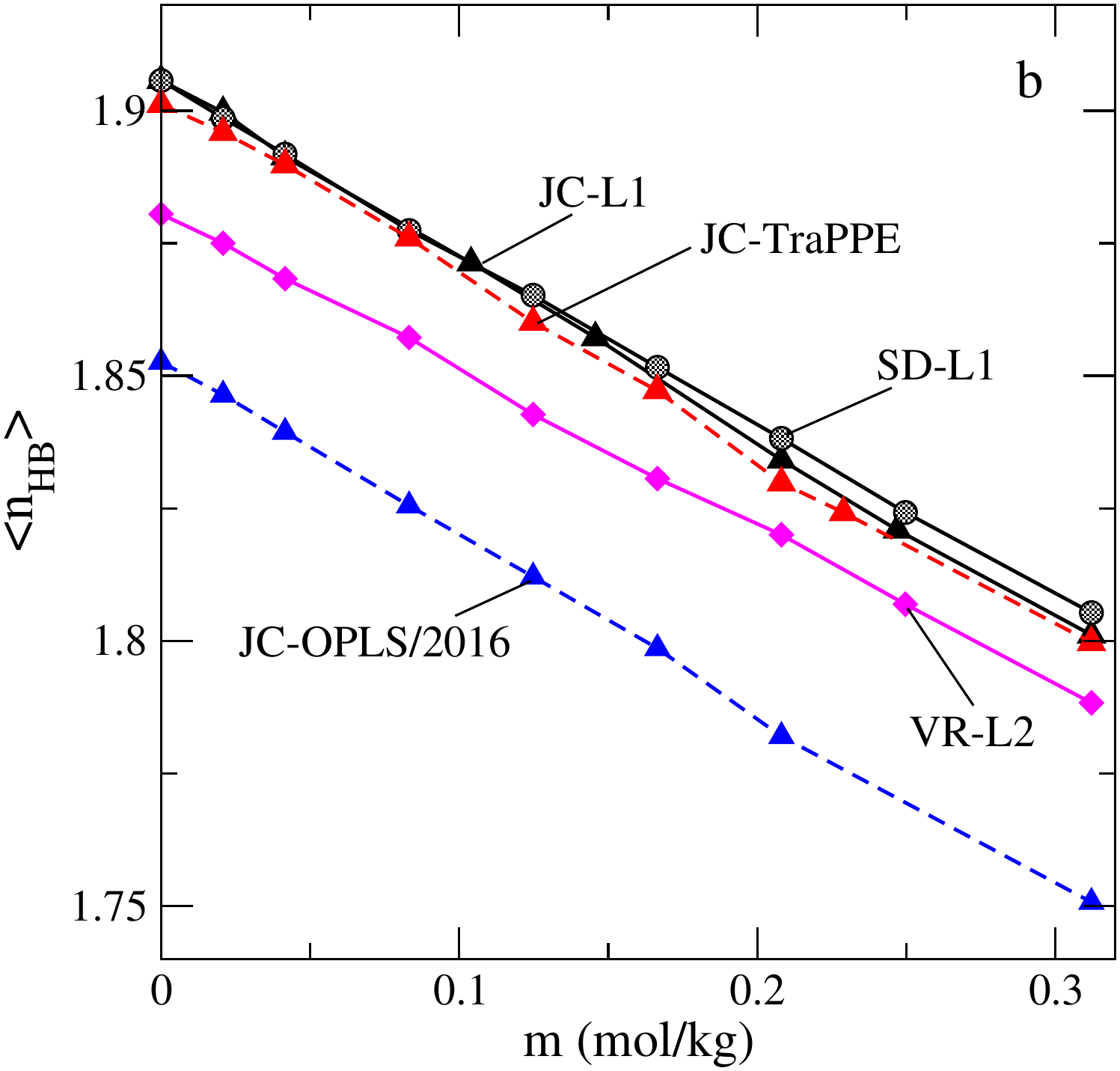}
\caption{(Colour online) 
Panel (a): Illustration of the change of the methanol structure upon adding ions
for the JC-OPLS/2016 model. The pair distribution functions are shown for
$m=0$ and $m=0.2081$ ($N_\text{ip}=0$ and $N_\text{ip}=20$, respectively).
Panel (b): Changes of the average number of H-bonds per methanol molecule upon
adding the ions to the solution for various models.
}
\label{fig4}
\end{figure}

\begin{figure}[!t]
\centering
\includegraphics[width=6.4cm,clip]{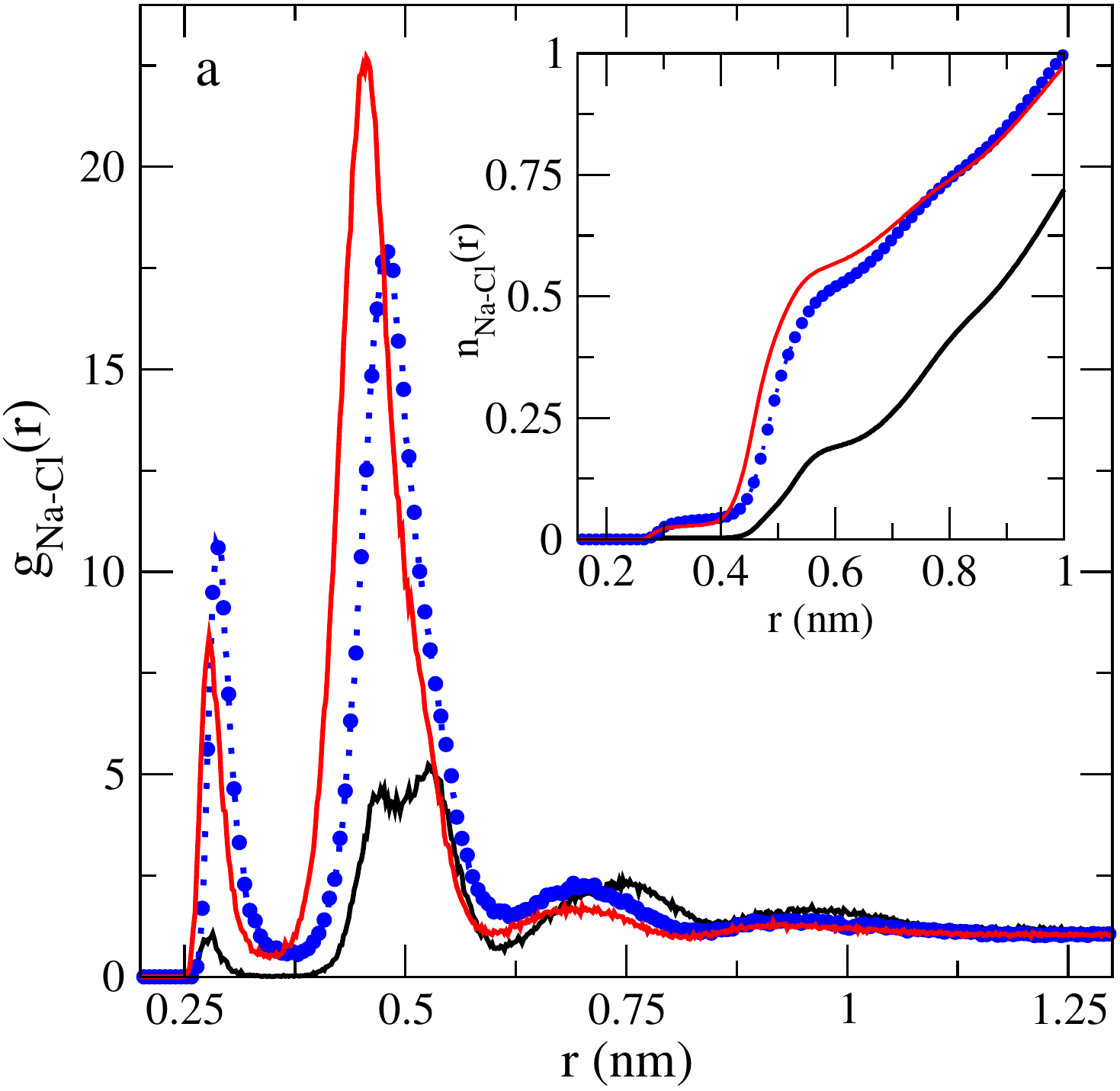}
\includegraphics[width=6.5cm,clip]{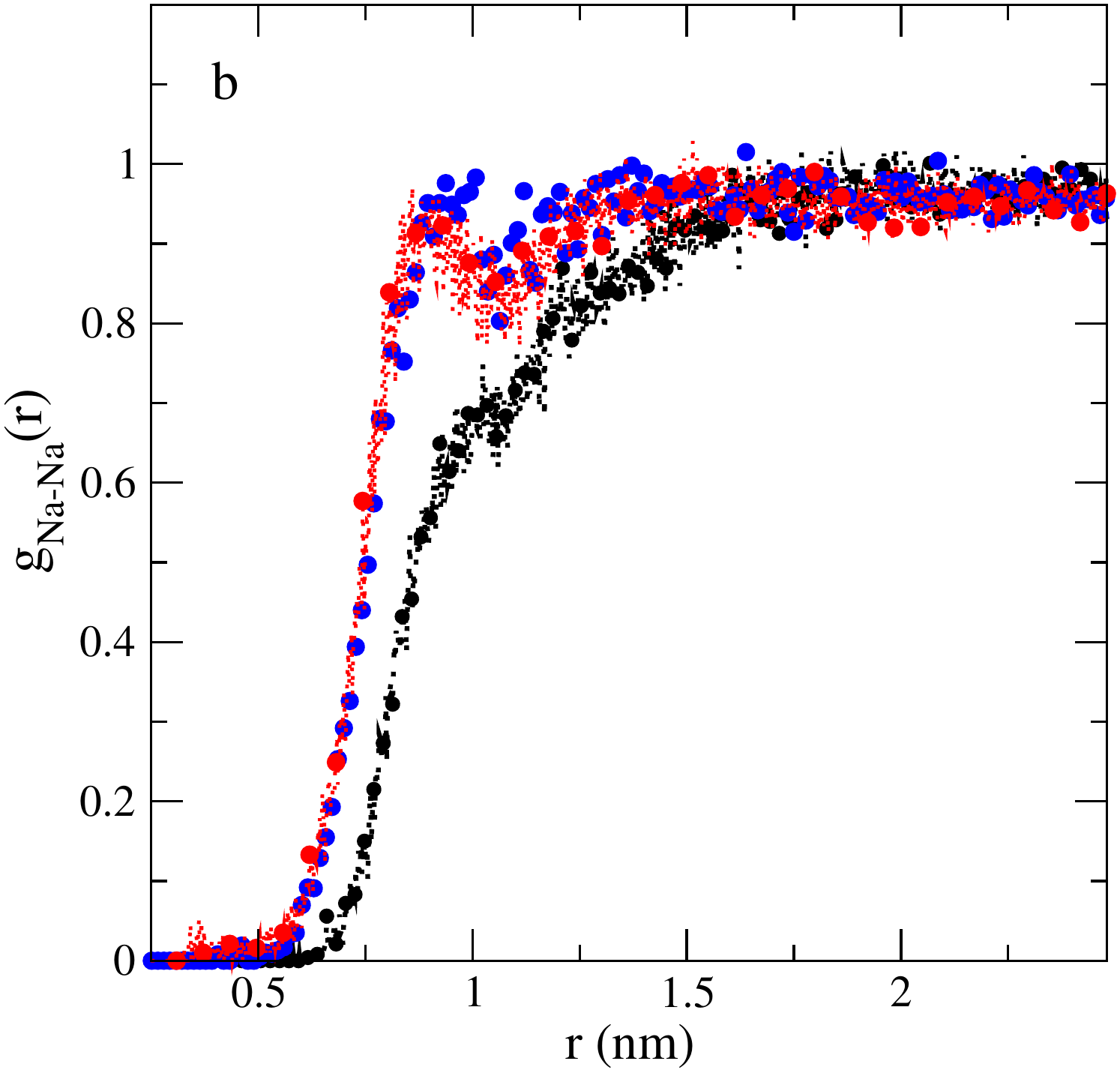}
\caption{(Colour online) 
Microscopic structure of NaCl methanol solutions following from the VR-L2, JC-TraPPE 
and JC-OPLS/2016 models in terms of ion-ion pair distribution functions at molality
$m=0.2081$ (black solid line --- OPLS/2016, red solid line --- TraPPE, blue
dotted line with circles --- L2 model, respectively). The inset to panel (a) shows
the running coordination number coming from $g_\text{Na--Cl}(r)$.
}
\label{fig5}
\end{figure}

Solvation of ions by methanol molecules perturbs the solvent structure and leads
to changes of the network of the hydrogen bonds. Hydrogen bonding in pure methanol
was explored on several occasions, see e.g.,~\cite{keith} for one of
the first discussions of the issue. In order to describe this effect we
have compared the PDFs providing the distributions of oxygens and hydrogens in pure
methanol and at a high molality, $m=0.2081$. Apparently, small changes of these distributions
in the solvent [figure~\ref{fig4}~(a)] result in a more pronounced changes of the average
number of hydrogen bonds [figure~\ref{fig4}~(b)]. Geometric, two-parameter (distance-angle) 
criterion, incorporated in GROMACS software was used to construct the
plot in figure~\ref{fig4}~(b). However, the position of the minimum of $g_\text{OO}(r)$ was
used in calculations as the donor-acceptor distance. We have already seen
that the height of the first maxima for O--O and O--H distribution functions
in pure methanol is smaller within the JC-OPLS/2016 model, in comparison 
with JC-TraPPe and VR-L2 models [figure~\ref{fig3}~(a)]. Consequently, the average number of H-bonds 
is lower in the former model compared to the other two, figure~\ref{fig4}~(b). This trend
preserves in the entire interval of molality studied in the present work.
As expected, the $\langle n_\text{HB}\rangle$ decreases with an increasing molality, because the 
solvation of ions restricts the configurations that satisfy the bonding criteria.
Alternatively to the geometric interpretation, the hydrogen bonding can be explored
in fine details using energetic criteria, see e.g.,~\cite{bako}. 

Our final remarks w.r.t. the behaviour of the pair distribution functions concern
the distribution of solute ions in the system at a high molality, $m=0.2081$.
Actually, this value is close to the experimental solubility of NaCl salt in methanol.
The PDFs of ions are shown in figure~\ref{fig5}.
The shape of the Na--Cl distribution is similar for three models in question,
the JC-TraPPE, VR-L2 and JC-OPLS/2016. All of them describe a low probability to find
contact ion pairs (CIP) and a higher probability to observe solvent-separated ion
pairs (SSIP), see panel a of figure~\ref{fig5}. Apparently, the results coming from the 
JC-TraPPE and VR-L2 models are similar. The VR-L2 model yields a stronger tendency
to find CIP compared to JC-TraPPE. By contrast, the JC-TraPPE predicts a higher
probability to find SSIP in comparison with VR-L2 model. The JC-OPLS/2016 behaves
very differently in quantitative terms, compared to JC-TraPPE and VR-L2. Namely,
the maxima describing CIP and SSIP are very much smaller evidencing a quite low
probability of ions pairing in this model. The behaviour of the corresponding 
running coordination numbers is described in the inset to panel a of figure~\ref{fig5}. 
It is important to mention that the maxima of $g_\text{Na--Cl}(r)$ for all models in question
are very much higher, in comparison to the aqueous solution of this salt, even
close to saturation, cf. figure~4 of \cite{benavides2} for the recently designed
so-called Madrid model. Principally, this is the result of strong electrostatic
attraction between ions in methanolic solutions due to a low dielectric constant
of methanol compared to water. Still, the fraction of CIP is much smaller than that
of SSIP, as it follows from the coordination numbers, indicating competition of 
solvation of ions with the direct Coulomb attraction.
The pair distribution functions of the like ions, see figure~\ref{fig5}~(b), slowly tend to unity reaching
it beyond $r \approx 0.75$ for VR-L2 and JC-TraPPE models. The growth of $g_\text{Na--Na}(r)$
is even much slower within the JC-OPLS/2016 model. This behaviour results from the
assumed value of molality and from strong electrostatic repulsion. There is no evidence for 
anti-phase oscillations or alternating charge distribution for ions and,  consequently,
for the formation of ion clusters involving more ions than pairs in these 
solutions. Similar observations concerning the absence of anti-phase
oscillations were discussed in \cite{benavides2}, while describing
Madrid model for NaCl aqueous solutions close to saturation limit.

To summarize the above discussion, we would like to note that the models
illustrated  in terms of pair distribution functions lead to 
qualitatively correct and physically well grounded conclusions concerning the
microscopic structure, in accordance with the previous developments of other authors
using computer simulation methods and experimental observations.
However, any kind of improvement could be attempted, if the experimental 
structure factors of such complex systems were  available,
then fitting of the computer simulation results to, e.g., diffraction experiments 
data could be performed along the lines proposed in \cite{pusztai2,pusztai3} for aqueous 
electrolyte solutions.
\newpage
\subsection{Self-diffusion coefficients of methanol and ions}

Evolution of the microscopic structure of solutions with molality results in 
the changes of the dynamic properties. We restrict our attention solely to the
evaluation of the self-diffusion coefficients of the species.
If the number of particles is sufficiently  big to provide good statistics,
it is reasonable to obtain the self-diffusion coefficient 
from the mean-square displacement (MSD) of a particle via Einstein relation,
\begin{equation}
D_{\alpha} =\frac{1}{6} \lim_{t \rightarrow \infty} \frac{\rd}{\rd t} \langle\vert {\bf 
r}_i^{\alpha}(\tau+t)-{\bf r}_i^{\alpha}(\tau)\vert ^2\rangle_{i,\tau}\,,
\end{equation}
where  $\alpha$ is the species index, and the average is taken over all particles 
(indexed by $i$) and time origins $\tau$.
To begin with, we used this procedure to obtain, $D_\text{met}$.
Default settings of GROMACS were used for the separation of the time origins. 
Moreover, the  fitting interval was chosen to be from 10\% to less than 50\% 
of the entire trajectory (with time extension not less than 100~ns).
The finite size correction has not been taken into account, however.

\begin{figure}[!b]
\centering
\includegraphics[width=6.5cm,clip]{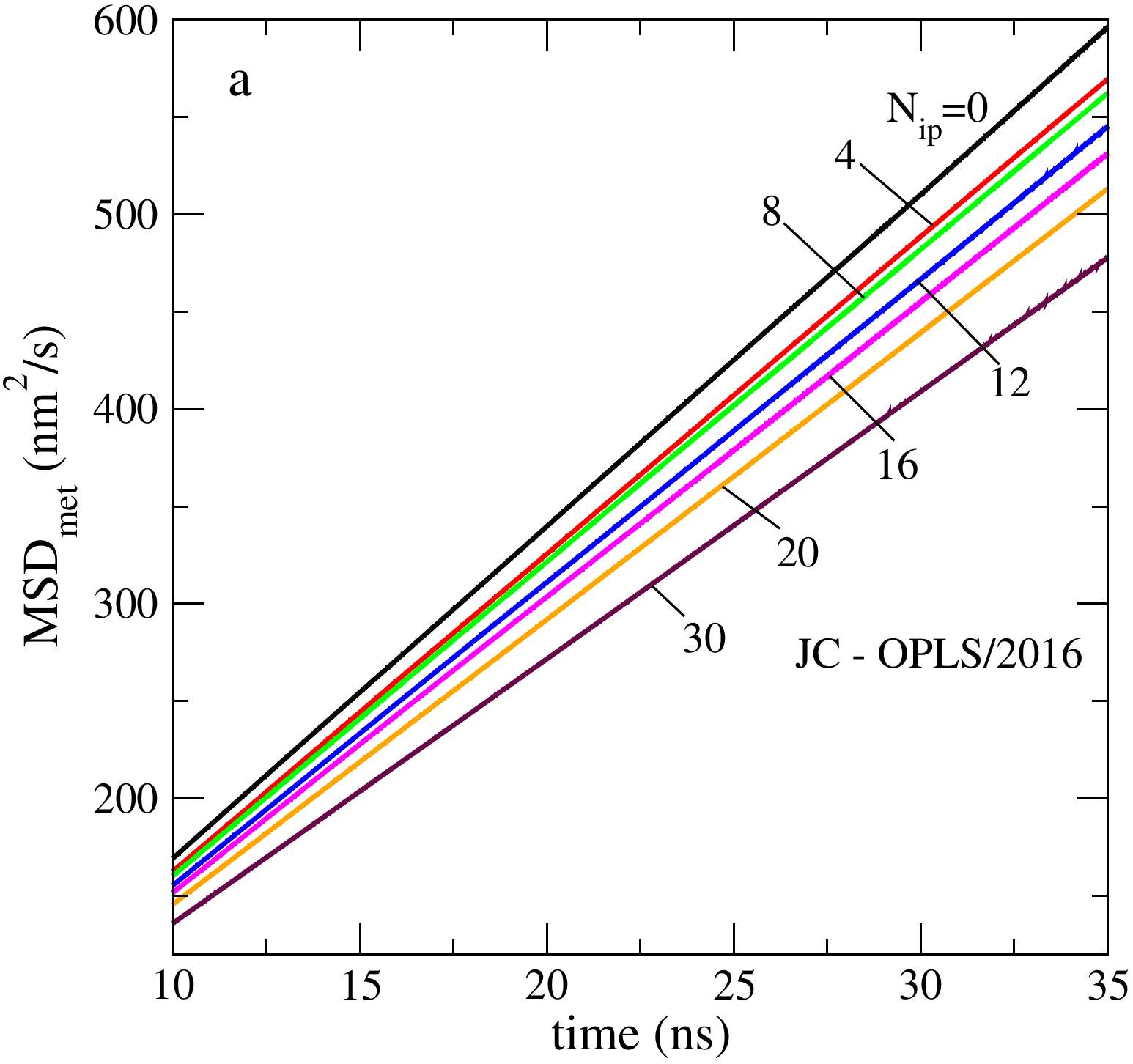}
\includegraphics[width=6.2cm,clip]{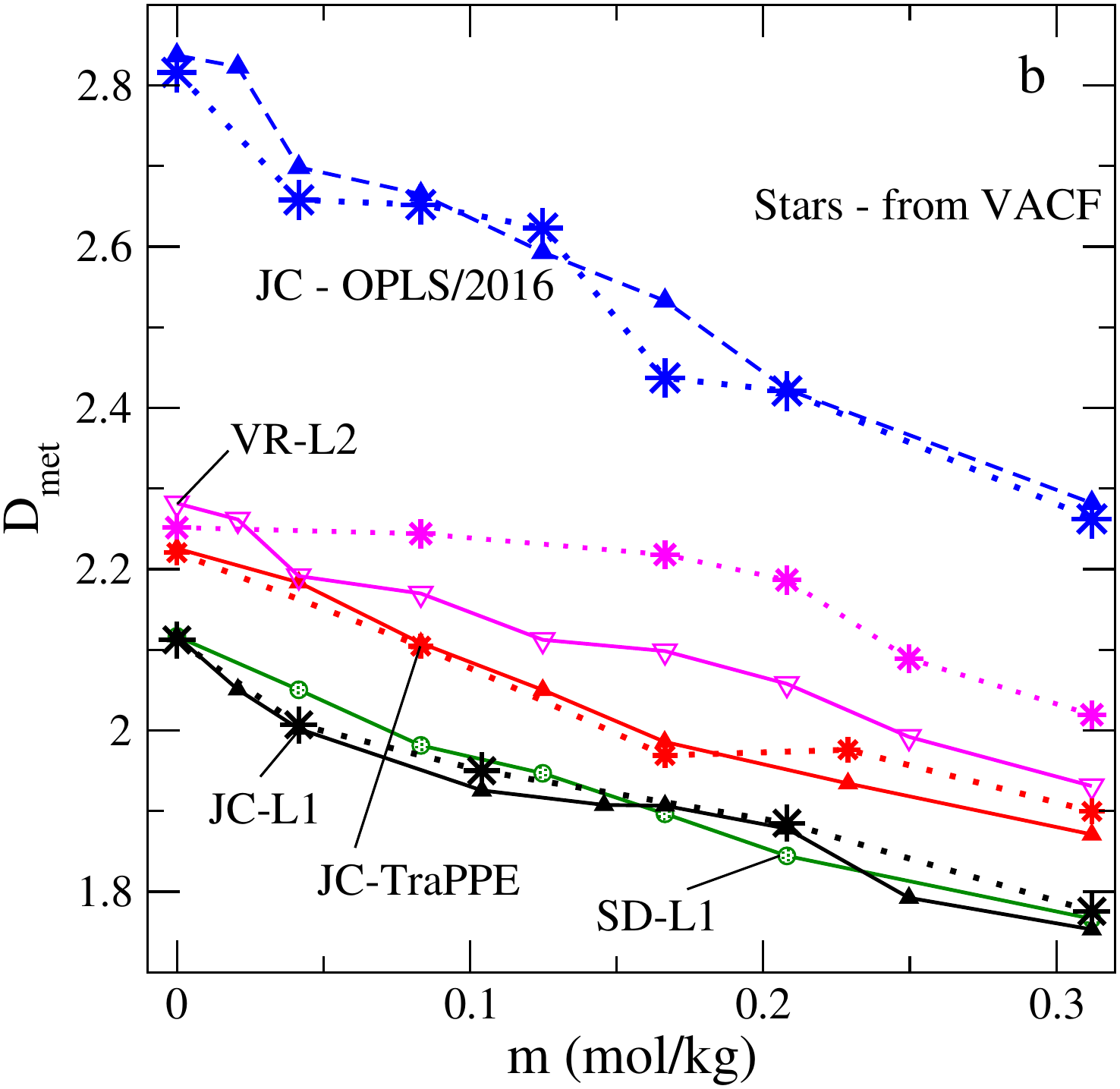}
\caption{(Colour online) 
Panel (a): Mean square displacement of methanol molecules with time in pure methanol and NaCl solutions
with different molality for the JC-OPLS2016 model. The number of ion pairs, $N_\text{ip}$, is indicated in the figure.
Panel (b): Changes of the methanol solvent self-diffusion coefficient, $D_\text{met}$,  on the NaCl solutions 
molality for different models using Einstein relation (lines and symbols) and from the Green-Kubo equation
(stars joined by dotted lines).
}
\label{fig6}
\end{figure}

\begin{figure}[!t]
\centering
\includegraphics[width=6.5cm,clip]{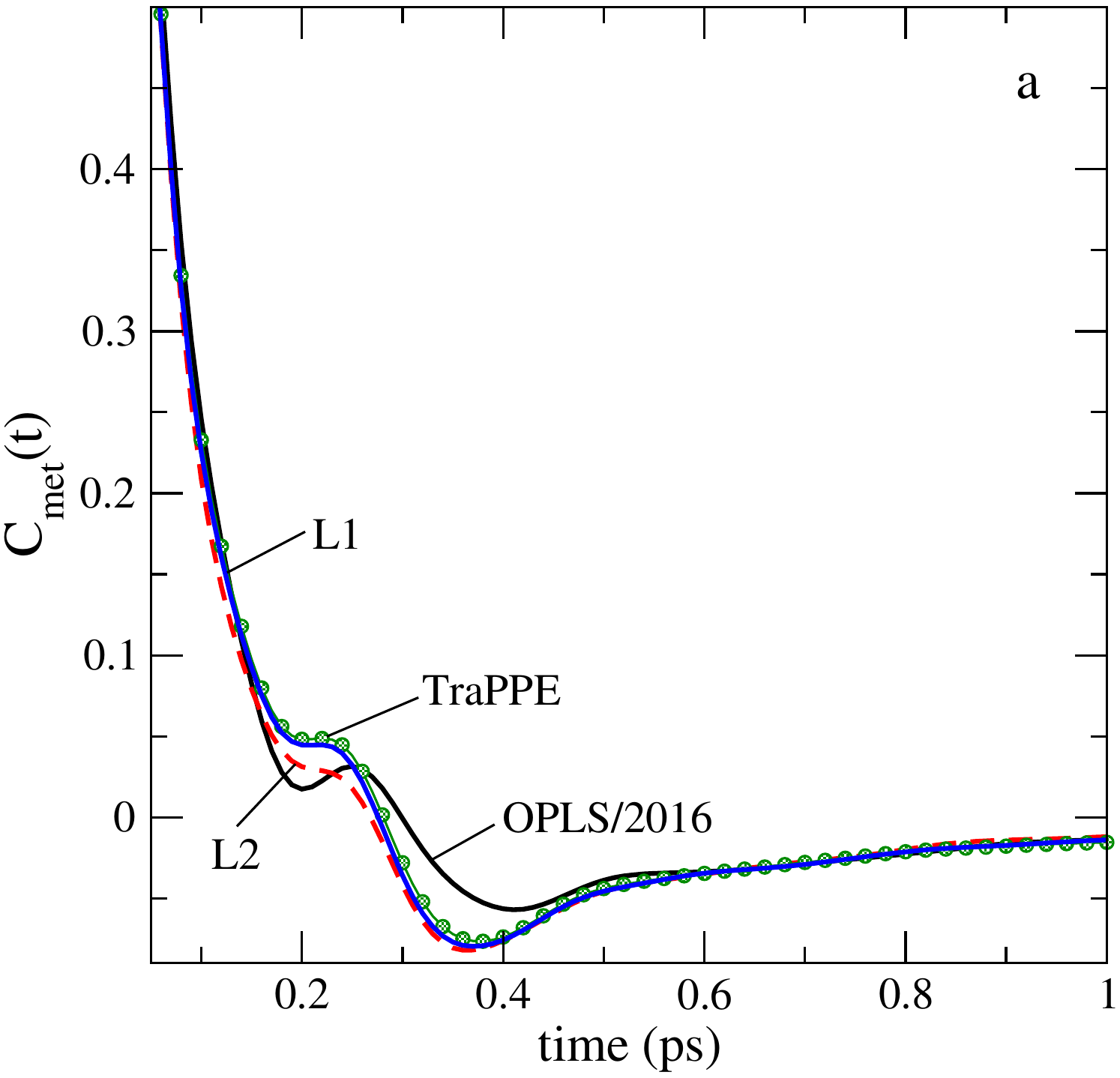}
\includegraphics[width=6.3cm,clip]{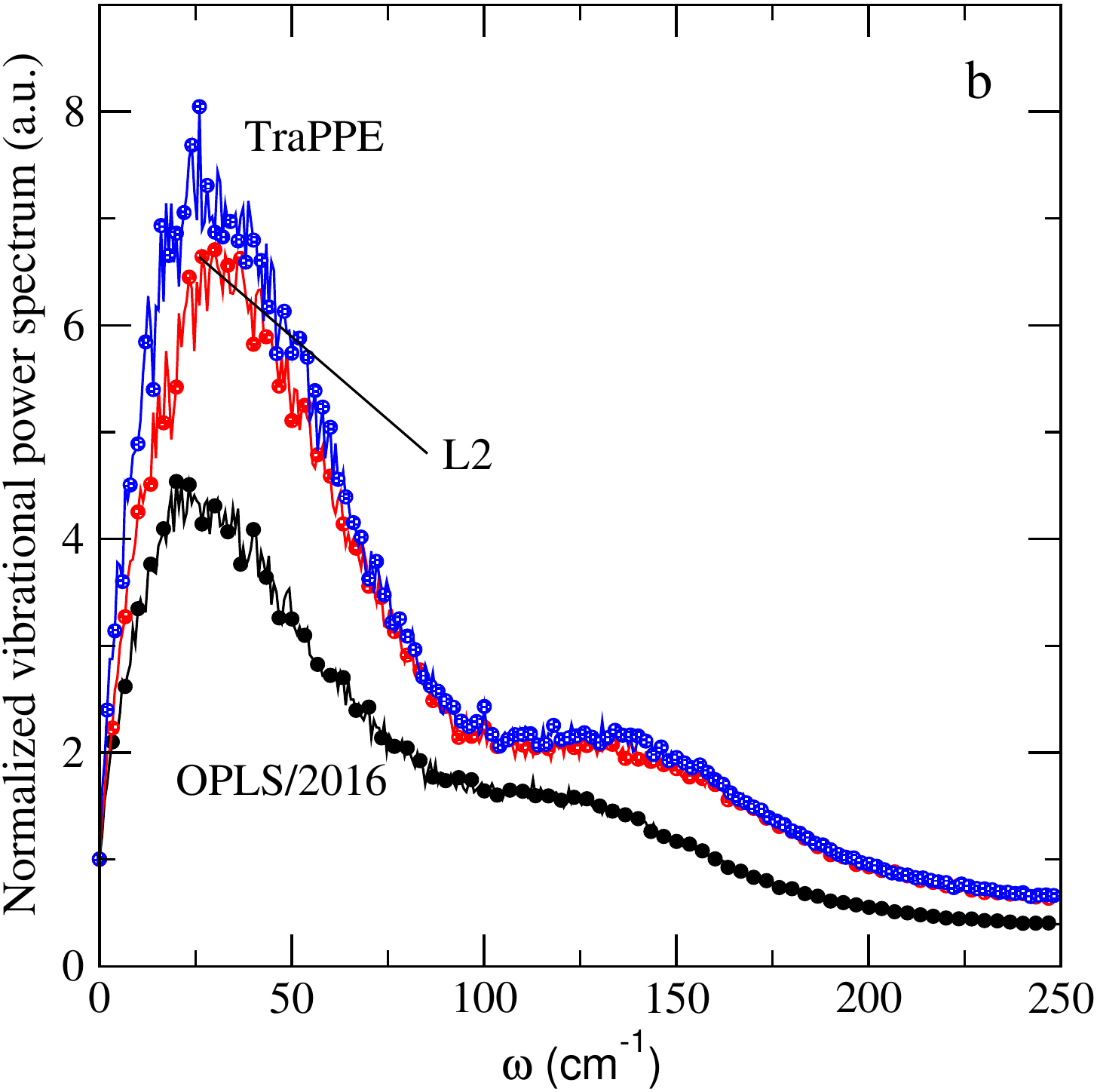}
\includegraphics[width=6.5cm,clip]{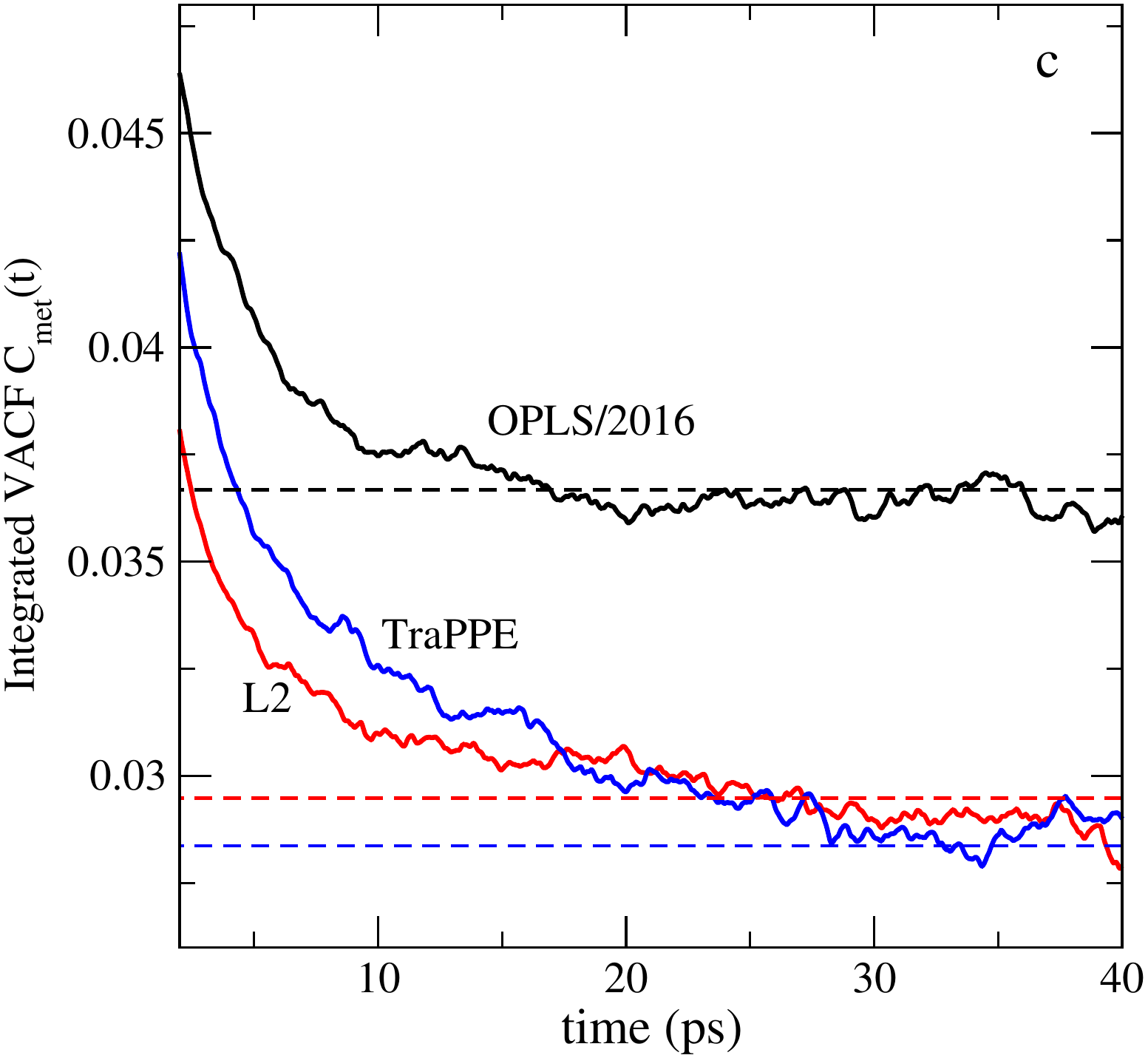}
\caption{(Colour online) 
Panel (a): Normalized velocity autocorrelation function of the COM of methanol molecules with different models.
Panel (b): Normalized vibrational power spectrum of methanol with different models.
Panel (c): Integrated normalized VACF of methanol with different models used for the evaluation of the
self-diffusion coefficient. Horizontal lines correspond to the results from MSD procedure.
}
\label{fig7}
\end{figure}

The dependence of MSD on time for systems at different molality, $m$,
for a single model (JC-OPLS/2016) is shown in figure~\ref{fig6}~(a). Each MSD line 
in figure~\ref{fig6}~(a) looks perfectly linear, such that the error margin at different $m$, is small. 
The entire set of data for different models is given in figure~\ref{fig6}~(b). 
The initial points of curves in figure~\ref{fig6}~(b) correspond to pure methanol. Our data
agree with the previously reported results, see e.g., table VII in \cite{vega} and
\cite{galicia2,guevara2}. The experimental result for pure 
methanol is at $2.44\times 10^{-5}$~cm$^2$/s.
The curves definitely predict a slowing down of methanol molecules in solutions with an increasing molality.
It is difficult to establish how good is the inclination of each curve
without experimental data that are unavailable, up to our best knowledge.
On the other hand, we are unaware of a similar set of data from molecular dynamics
simulations, in contrast to, for example, figure~13 of \cite{benavides2} for
aqueous solutions of NaCl. 

In order to get more confidence into the quality of the results obtained, we have
attempted to perform calculations of $D_\text{met}$ by an alternative method.
Namely, the translational self-diffusion coefficients can be determined from the velocity
auto-correlation functions (VACFs) using the Green-Kubo equation,

\begin{equation}
D_{\alpha} = \frac{k_\text{B}T}{m_\alpha} \int_{0}^{\infty} C_{\alpha}^{vv}(t) \rd t,
\end{equation}
where $C_{\alpha}^{vv}(t)$ is the normalized velocity autocorrelation function of species $\alpha$,
\begin{equation}
C_{\alpha}^{vv}(t) = \langle v_i^{\alpha}(\tau+t) v_i^{\alpha}(\tau)\rangle_{i,\tau}/\langle v_i^{\alpha}(\tau)^2\rangle_{i,\tau}\,,
\end{equation}
where the average is taken over all particles indexed by $i$ and time origins $\tau$,
similar to Einstein formula.  To realize the entire procedure, first, we have used the final configuration file
from the $NPT$ simulations and performed an additional $NVT$ run for each model with the duration of  $100$~ps, 
the time step was chosen at $1$~fs.  The normalized VACF for methanol center of mass, 
following from the GROMACS utility, are plotted in figure~\ref{fig7}~(a).
It appears that the results for L1 and TraPPE models are very similar. On the other hand,
the results for L2 just differ from the ones for L1 and TraPPE models by the shape of a shoulder.
Finally, the VACF describing the OPLS/2016 model differ from the other curves by the shape of a shoulder,
by the depth and position of the principal minimum. An overall shape of the VACF reported here
favourably agrees with the result given in figure~4 of \cite{sese}. On the other hand,
the PHH model for methanol is characterized by a slightly less prominent shoulder, cf. figure~13 of
\cite{marx}. 

The Fourier transform of the VACF yields the vibrational power spectrum
plotted in figure~\ref{fig7}~(b). Two relevant bands of the spectrum were discussed for methanol
previously~\cite{guevara2}. Namely, the band at frequency around  50~cm$^{-1}$ and a shoulder at 
frequency around 150~cm$^{-1}$. The first maximum can be assigned to the motion of the particles inside the cage 
formed by their neighbors whereas the shoulder is attributed to the presence of 
hydrogen bonds between methanol molecules. From the inspection of the spectrum calculated for
different models we see that the shoulder is less pronounced for the OPLS/2016 model,
which is due to a lower average number of H-bonds per molecule, in comparison with the
other models of methanol, cf. figure~\ref{fig4}~(b).  

The knowledge of the VACF permits to obtain self-diffusion coefficients of the species.
The integral entering Green-Kubo equation should be evaluated as precisely as possible,
because it is further multiplied by a big number, $k_\text{B}T/m_{\alpha}$ for each species. 
In the case of methanol molecules, we have a good guide to do that using the MSD results.
In figure~\ref{fig7}~(c), we plot the horizontal lines resulting from the MSD data for each model that should
correspond to the integrated normalized VACF. The curves in that figure result from
the VACF integration. One needs to find a plateau region, estimate the integral and obtain
$D_\text{met}$. The plateau is located at different time intervals for different models, as
intuitively expected.
This kind of procedure has been realized for a set of
desired molality points for each model. The results are given in figure~\ref{fig6}~(b) as stars joined by dotted lines. 
In the case of pure methanol, two methods produce quite similar results 
and the margin of error is less than 2\%.
On the other
hand, trends of slowing down self-diffusion of methanol molecules are reproduced as well.
The best agreement between the data resulting from two methods has been obtained for
JC-L1 and JC-TraPPE models. It is also quite satisfactory for the
JC-OPLS/2016 model. The inaccuracy of data slightly increases with an increasing 
ion concentration, the margin of error becomes of the order of 3--4\% at the highest concentration
studied. The MSD results are apparently more confiable.

Concerning the evaluation of the self-diffusion coefficients of Na$^+$ and Cl$^-$ ions 
in methanolic solutions, we would like to make the following comments. 
We are aware of only scarce data for this specific system. By contrast, there were
several reports that describe the dependence of self-diffusion coefficients of ions
on molality for aqueous solutions, see e.g., \cite{benavides2,laaksonen2,hartkamp,bouazizi2}.
Both methods, i.e., the Einstein relation and Green-Kubo equation, were used
in above cited works.

\begin{figure}[!t]
\centering
\includegraphics[width=6.8cm,clip]{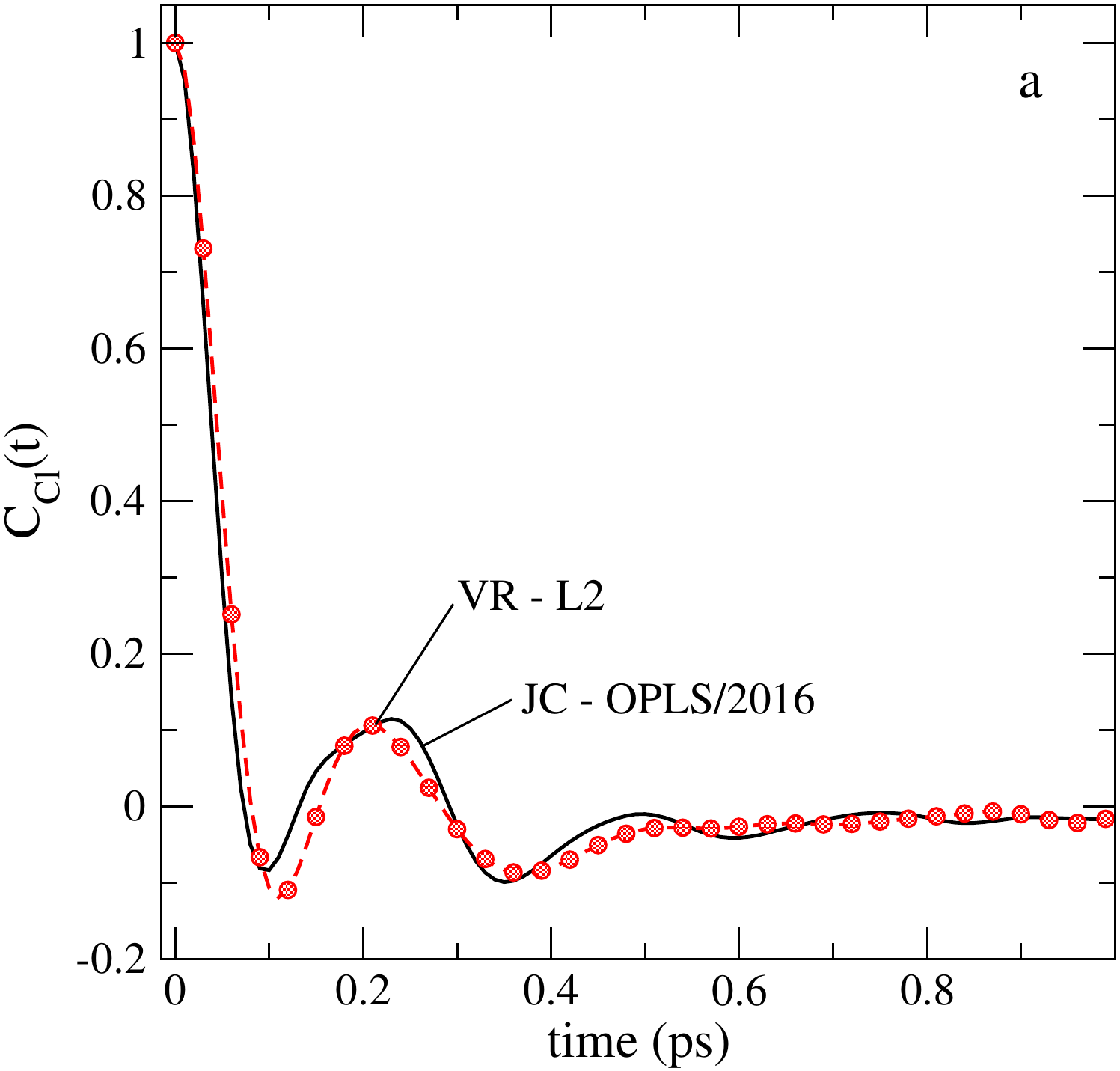}
\includegraphics[width=6.8cm,clip]{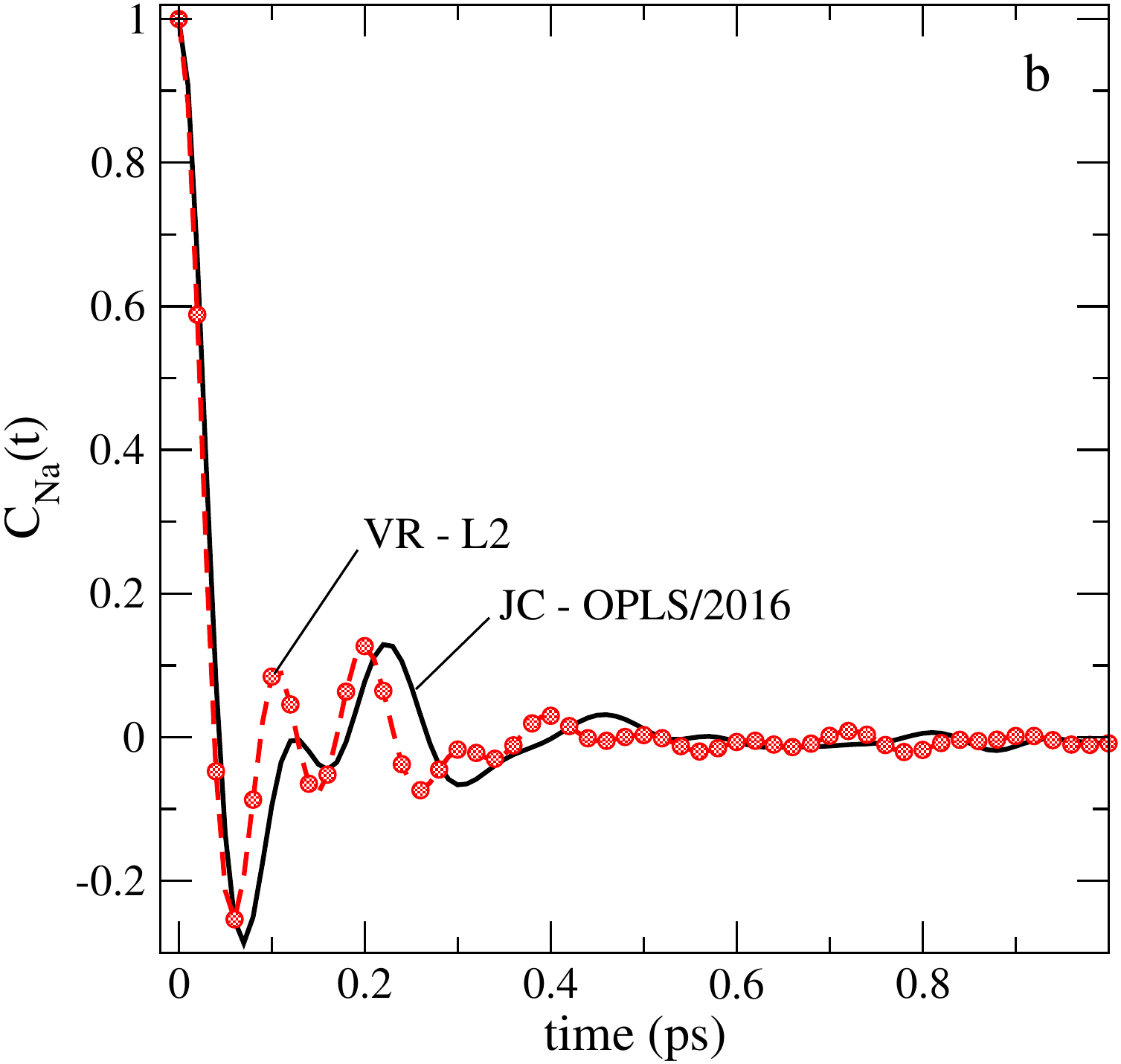}
\caption{(Colour online) 
Panels (a) and (b): Normalized velocity autocorrelation function of Cl anion and Na cation in the
methanolic solution at molality 0.0416 ($N_\text{ip}=4$) from the JC-OPLS/2016 and
VR-L2 models, respectively. 
}
\label{fig8}
\end{figure}

Recall, that the number of ion pairs, $N_\text{ip}$, in the studied systems is quite low. 
Therefore, using the mean square
displacement procedure on a long time scale leads to an augmenting uncertainty due to
sampling statistics, as discussed for example in~\cite{laaksonen2}. In close similarly to that
work, we took the slope of the MSD in the interval between 100 and 300~ps to obtain rough 
estimates for $D_\text{Cl}$ and $D_\text{Na}$ through the Einstein relation. The results are collected
in table~\ref{tab3}. On the other hand, the Green-Kubo equation was used. The VACFs for ions for
two models in question in solutions at a rather low molality are shown in figure~\ref{fig8} for 
illustrative purposes. The curves agree well with the ones reported in figure~2 of~\cite{sese}. The PHH model for methanol is characterized by a slightly different shape of the
VACF for Na$^+$, cf. figure~\cite{marx}. The values for $D_\text{Cl}$ and $D_\text{Na}$ resulting
from the VACFs route at a low and at a high molality are given in table~\ref{tab3} as well.
Apparently, the agreement between the data coming from two procedures is less pronounced
for the JC-OPLS/2016 model. General trend of behaviour of $D_\text{Cl}$ and $D_\text{Na}$ is that
they decrease in magnitude with augmenting molality in qualitative similarity to what 
was observed for aqueous solutions, see e.g., figure~14 of \cite{benavides2}. Our
data in table~\ref{tab3} for VR-L2 differ but do not contradict the results for NaCl 
salt in methanol from the report of \cite{reiser} at molality $m \approx0.0313$.
The self-diffusion coefficient for Na$^+$ is lower than for Cl$^-$ according to
the predictions of all models studied.  


  \begin{table}[!b]
  \small
    \caption{ Self-diffusion coefficients of ions for the methanolic solution of NaCl
    at two different molalities using different force fields for ions and methanol.
    The results were obtained using Einstein relation and Green-Kubo equation, respectively.
    Technical details of calculations are given in the text.}
    \label{tab3}
   \vspace{2ex}
    \begin{center}
     \begin{tabular}{c| c c| c c| c c| c c| c c }
      \hline
       $m=0.0416$  &    \multicolumn{2}{c|}{JC-OPLS2016}  &  \multicolumn{2}{c|}{VR-L2}  &    \multicolumn{2}{c|}{JC-TraPPE} &  \multicolumn{2}{c|}{JC-L1}  &  \multicolumn{2}{c}{SD-L1}  \\
                 &    Cl &  Na  &  Cl &  Na  &  Cl &  Na &  Cl &  Na  &  Cl &  Na  \\
       \hline
       MSD  & 1.192 & 0.936 & 0.963 &0.910 & 0.965 &0.858 & 0.823 &0.744 & 0.992 &0.836 \\
       VACF & 1.231 & 0.857 & 1.084 &0.845 & 1.014 &0.944 & 1.013 &0.873 & 1.112 &1.067  \\
       \hline
       $m=0.2081$  &    \multicolumn{2}{c|}{JC-OPLS2016}  &  \multicolumn{2}{c|}{VR-L2}  &    \multicolumn{2}{c|}{JC-TraPPE} &  \multicolumn{2}{c|}{JC-L1}  &  \multicolumn{2}{c}{SD-L1}  \\
                 &    Cl &  Na  &  Cl &  Na  &  Cl &  Na &  Cl &  Na  &  Cl &  Na  \\
       \hline
       MSD  & 0.961 & 0.798 & 0.868& 0.758 & 0.838& 0.772 & 0.774 &0.744 & 0.775 &0.745 \\
       VACF & 1.035 & 0.748 & 0.979& 0.836 & 0.944 &0.839 & 0.874 &0.803 & 0.853 &0.744  \\
       \hline
  \end{tabular}
  \end{center}
\end{table}

\subsection{Solvent dielectric constant}

In this subsection we would like to comment on a few peculiar issues of the dielectric response 
of NaCl solutions with methanol solvent.
Since an electrolyte solution is a conductor, the mobile ions move in response to an
imposed static electric field. It is common to define a static equilibrium property that plays a
role similar to the dielectric constant of an insulating fluid, see, e.g., a comprehensive discussion of
the problem in~\cite{hubbard}. In several works 
that describe the application of molecular dynamics computer simulations to electrolyte solutions,
the counterpart of the dielectric constant is estimated in terms of the time-average of the fluctuations of the
total dipole moment of the system~\cite{caillol,zasetsky},
obtained by the summation of the dipole moment vectors of
the entire set of solvent molecules in the simulation box,

\begin{equation}
\varepsilon_\text{solvent}=1+\frac{4\piup}{3k_\text{B}TV}(\langle{\bf M}^2\rangle-\langle{\bf M}\rangle^2),
\end{equation}
where $V$ is the simulation cell volume.

\begin{figure}[!t]
\centering
\includegraphics[width=6.8cm,clip]{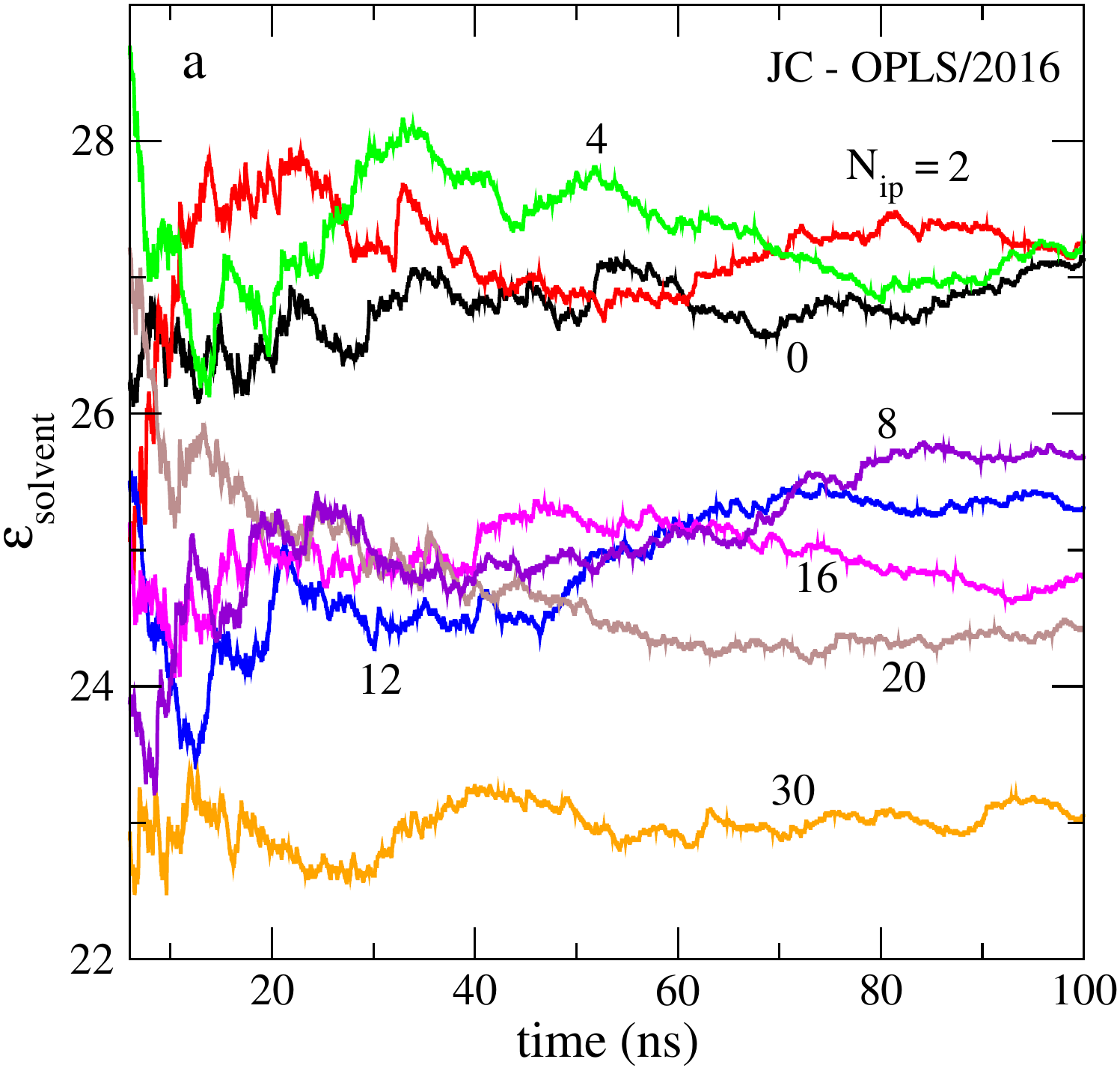}
\includegraphics[width=5.9cm,clip]{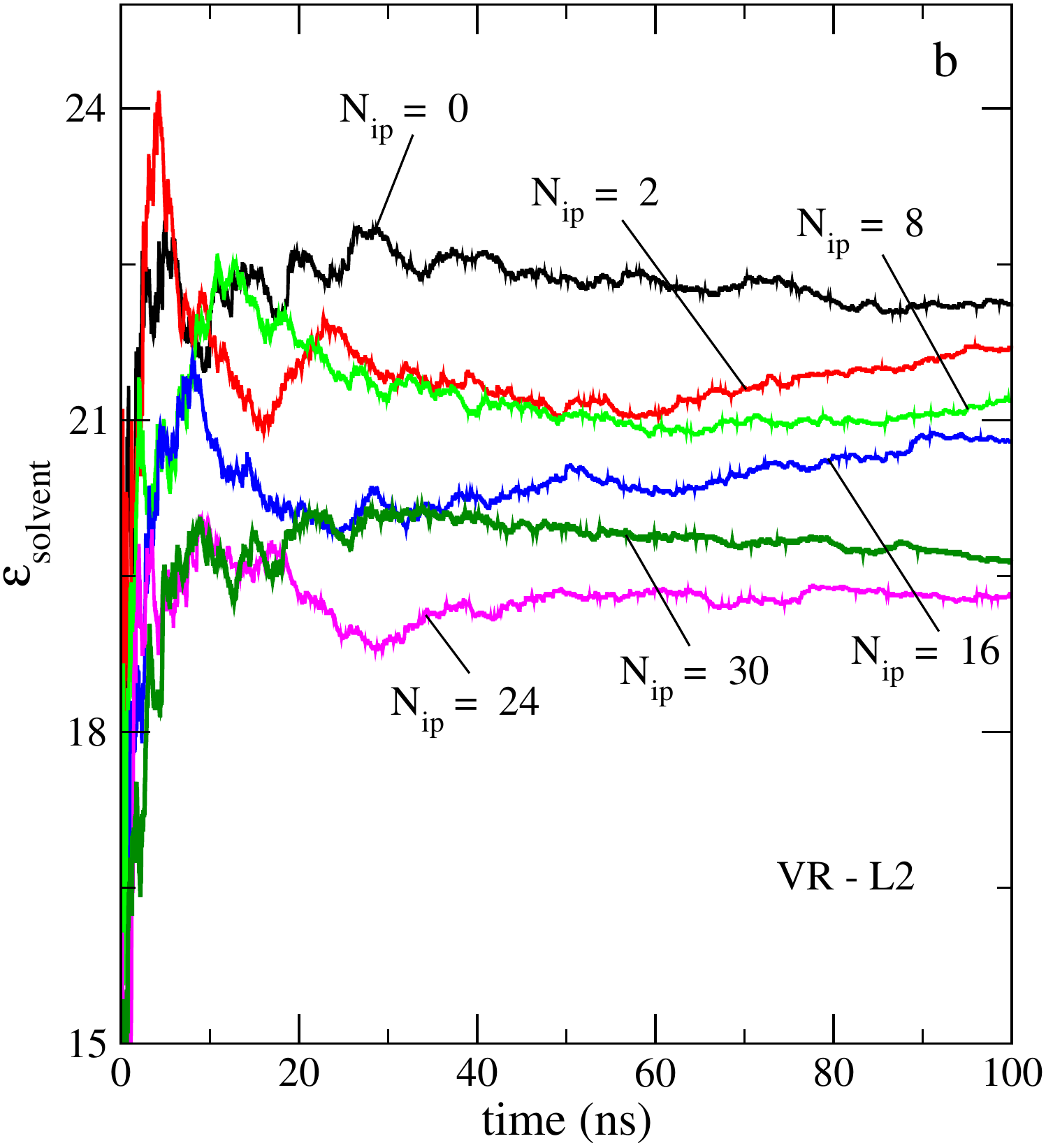}
\includegraphics[width=6.5cm,clip]{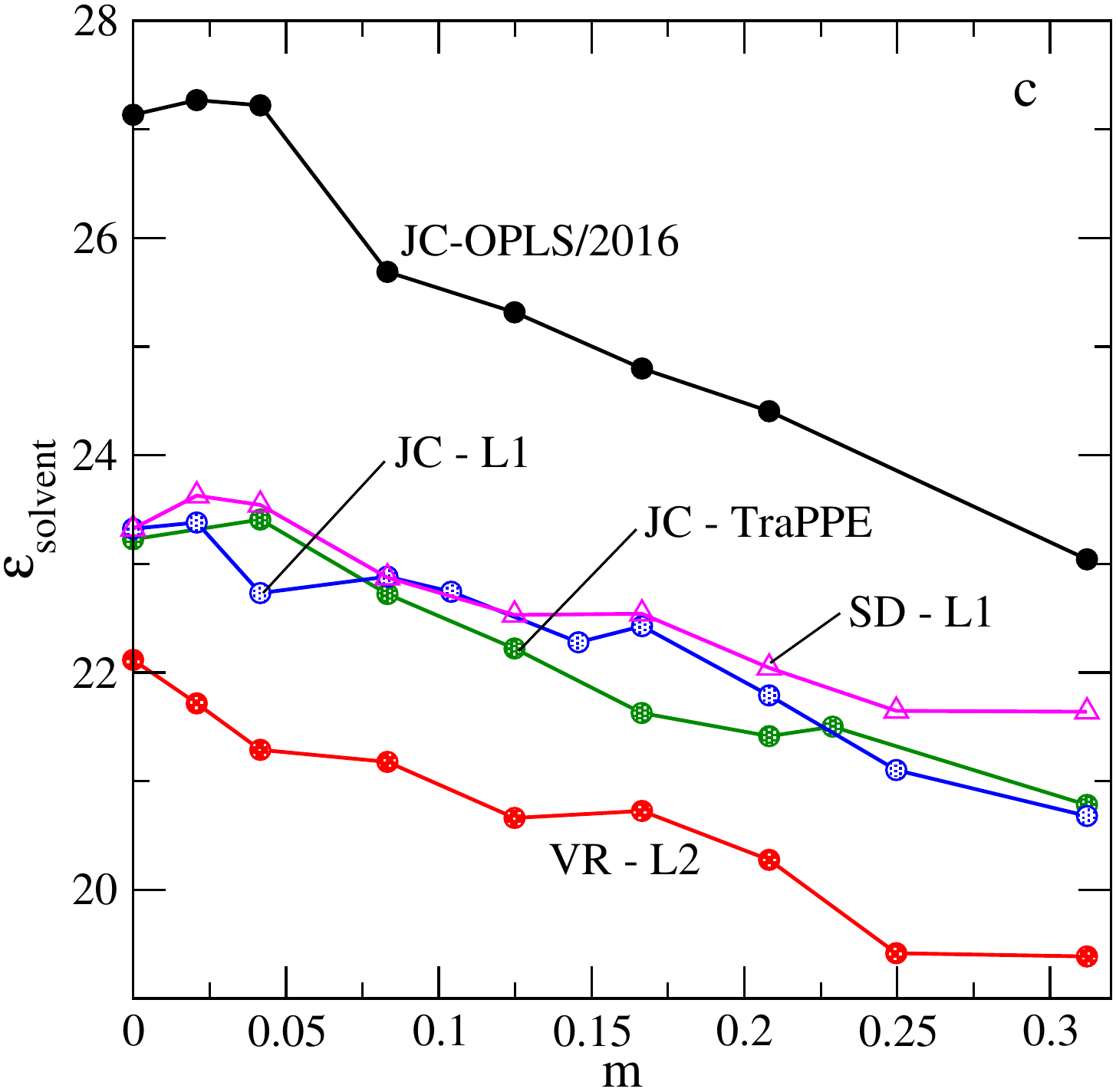}
\caption{(Colour online) 
Panel (a): The dependence of the solvent dielectric constant on time at different
amount of dissolved ions for the JC-OPLS2016 model.
Panel (b): The same as in panel (a), but for the VR-L2 model.
Panel (c): Illustration of the dielectric decrement, i.e., the dependence of
the methanol solvent static dielectric constant, $\varepsilon_\text{solvent}$, on the  NaCl solutions
molality (the models are indicated in the figure).
}
\label{fig9}
\end{figure}

As discussed in detail in \cite{barthel}
for aqueous electrolyte solutions with completely dissociated solute
molecules (NaCl in water is an example, see table 7.6 of~\cite{barthel}),
the static permittivity of the sample, $\varepsilon$,
decreases with increasing the electrolyte concentration. This permittivity represents
the static permittivity of the solvent in the solution. Moreover, the concentration dependent decrease of
the solvent permittivity defines the dielectric decrement. For aqueous solutions of NaCl
it was discussed using, for example, the micro-field approach~\cite{gavish} and 
the results are in favourable agreement with experimental data~\cite{buchner}.
We would like to briefly report this property as a function of ions concentration for different models
of solutions. Namely, in figures~\ref{fig9}~(a) and \ref{fig9}~(b), the time evolution of $\varepsilon_\text{solvent}$ 
at different concentrations of ions in a solution, is shown for the JC-OPLS/2016 and VR-L2 models, respectively.
It is known already that quite long MD runs are necessary to adequately describe the static dielectric
constant of pure insulating liquids.  Even after 100--120~ns
runs, the data in these two panels do not show convergence to a perfectly  well defined constant. 
Therefore, the averaging over the final time interval was performed.
However, it is important to mention that the static dielectric constant of methanol
predicted by the OPLS/2016 model is closer to the experimental value equal to 33.1, 
compared to all other models of this study. In addition, 
one can observe that the solvent dielectric constant decreases with an increasing ion concentration, figure~\ref{fig9}~(c). 
It is impossible to evaluate how good is the magnitude of the decrement of curves without experimental
data for this particular property. As it follows from the data in figure~\ref{fig9}~(c), the downward inclination
of the curves is less pronounced, in comparison with the recent report concerning the modelling of
the permittivity of electrolyte solutions~\cite{mollerup} (see figure~5 of this work 
for the dependence of $\varepsilon_\text{solvent}$ on molarity for various salts in methanol).

\subsection{Surface tension}
The simulations at surface tension calculations aiming at each point of composition axis were
performed by taking the final configuration of particles from the isobaric run.
Next, the box edge along $z$-axis was extended by a factor of 3, generating an elongated box with a liquid slab
and two liquid-mixture-vacuum interfaces in the $x{-}y$ plane, in close similarity to the procedure
used in~\cite{vanderspoel}.
The total number of methanol molecules, $3 \times 10^3$, is reasonable to yield an area of the
$x{-}y$ face of the liquid slab sufficiently big. The elongation of the liquid slab along
$z$-axis is satisfactory as well.
The executable molecular dynamics file was modified by deleting
a fixed pressure condition just preserving V-rescale thermostatting with the same parameters
as in $NPT$ runs. Other corrections were not employed.
The values for the surface tension, $\gamma$, follow from the combination of the time
averages for the components of the pressure tensor,
\begin{equation}
\gamma = \frac {1}{2} L_z \Big\langle\Big[P_{zz}-\frac{1}{2}(P_{xx}+P_{yy})\Big]\Big\rangle,
\end{equation}
where $P_{ij}$ are the components of the pressure tensor along $i,j$ axes, and  $\langle\ldots\rangle$
denotes the time average.

\begin{figure}[!t]
\centering
\includegraphics[width=7.0cm,clip]{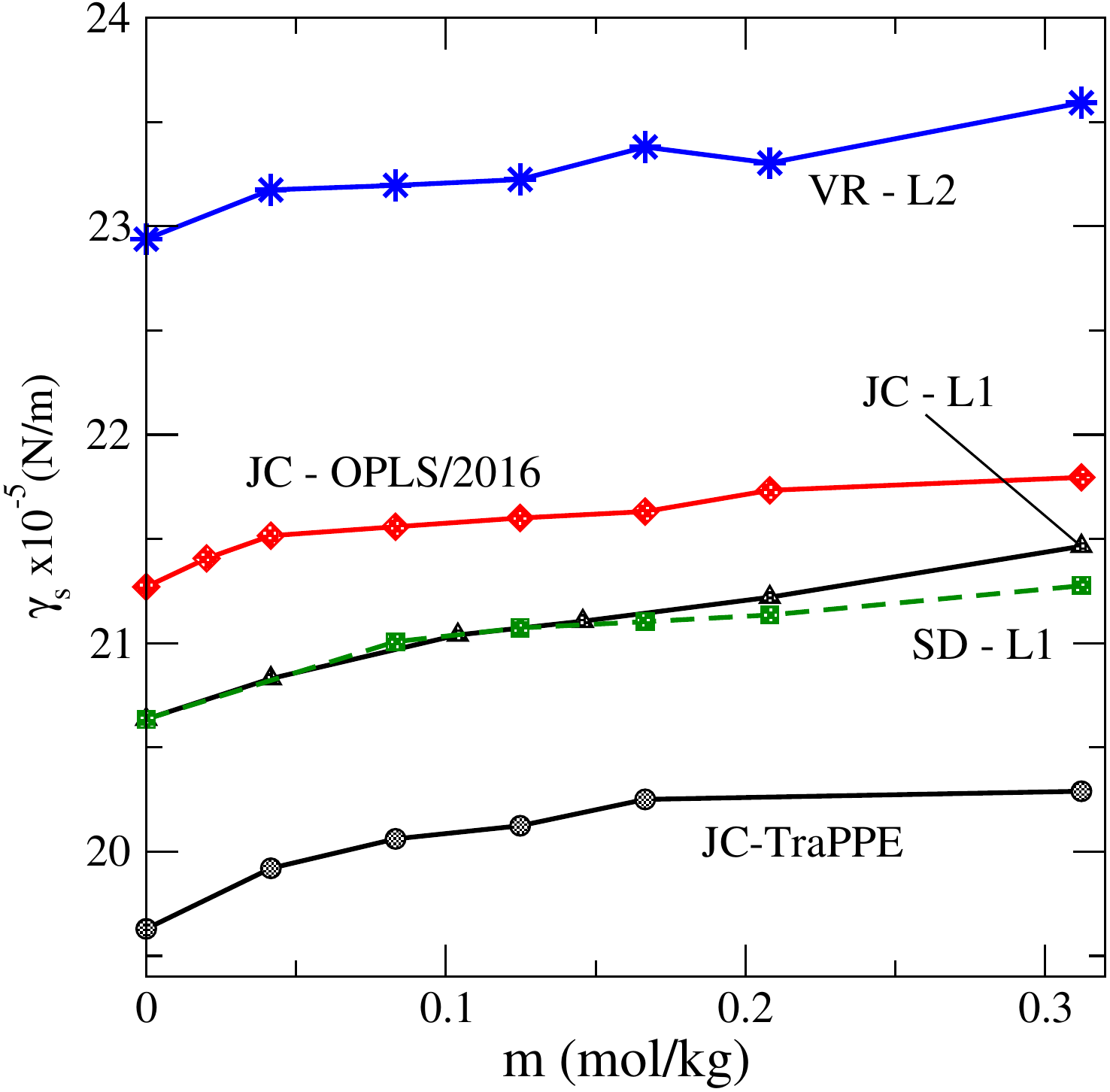}
\caption{(Colour online) 
Changes of the surface tension for the vapour-liquid interface of NaCl methanol solution with molality 
for different methanol models.
}
\label{fig10}
\end{figure}

\begin{figure}[!t]
\centering
\includegraphics[width=7.7cm,clip]{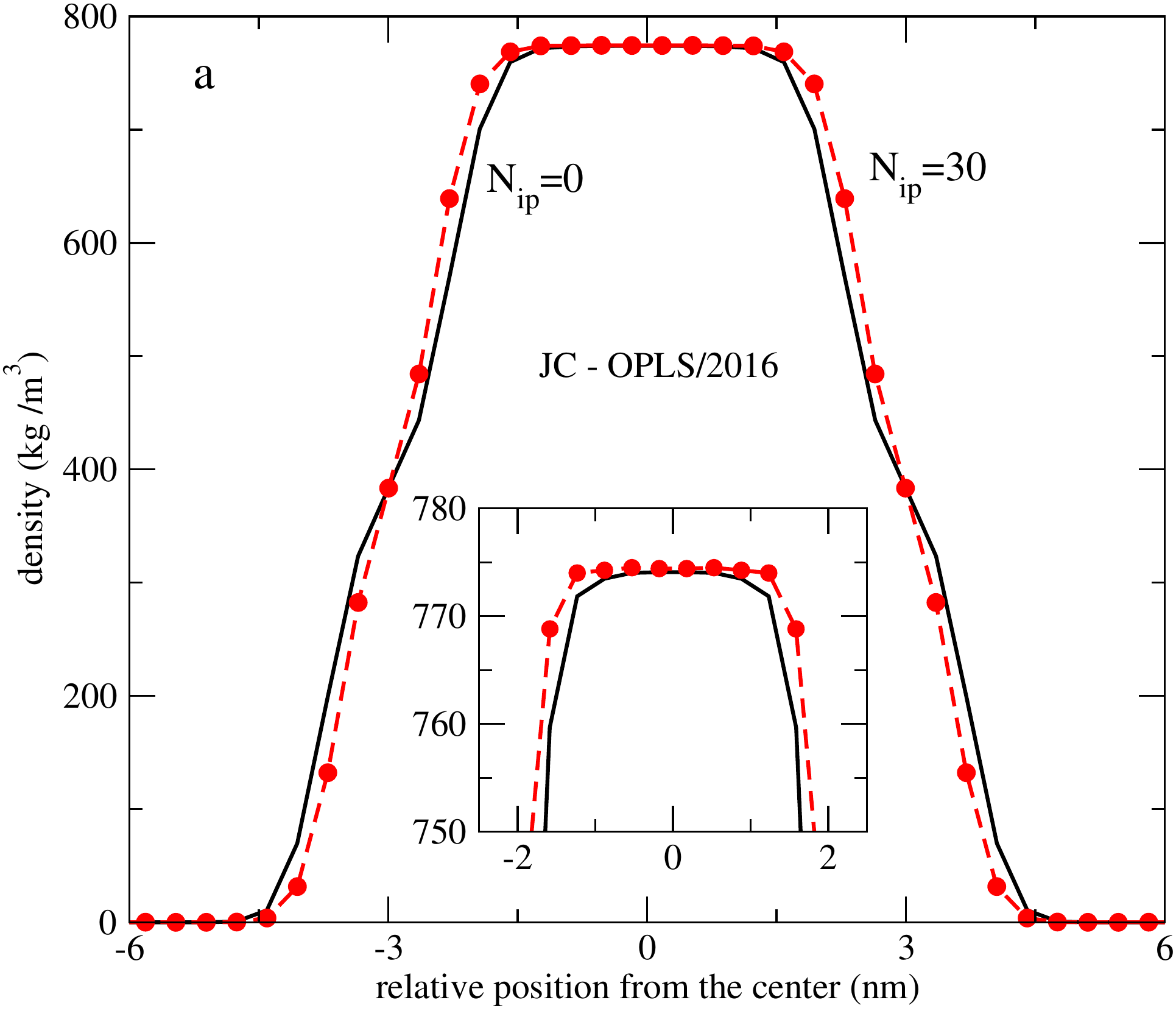}
\includegraphics[width=6.8cm,clip]{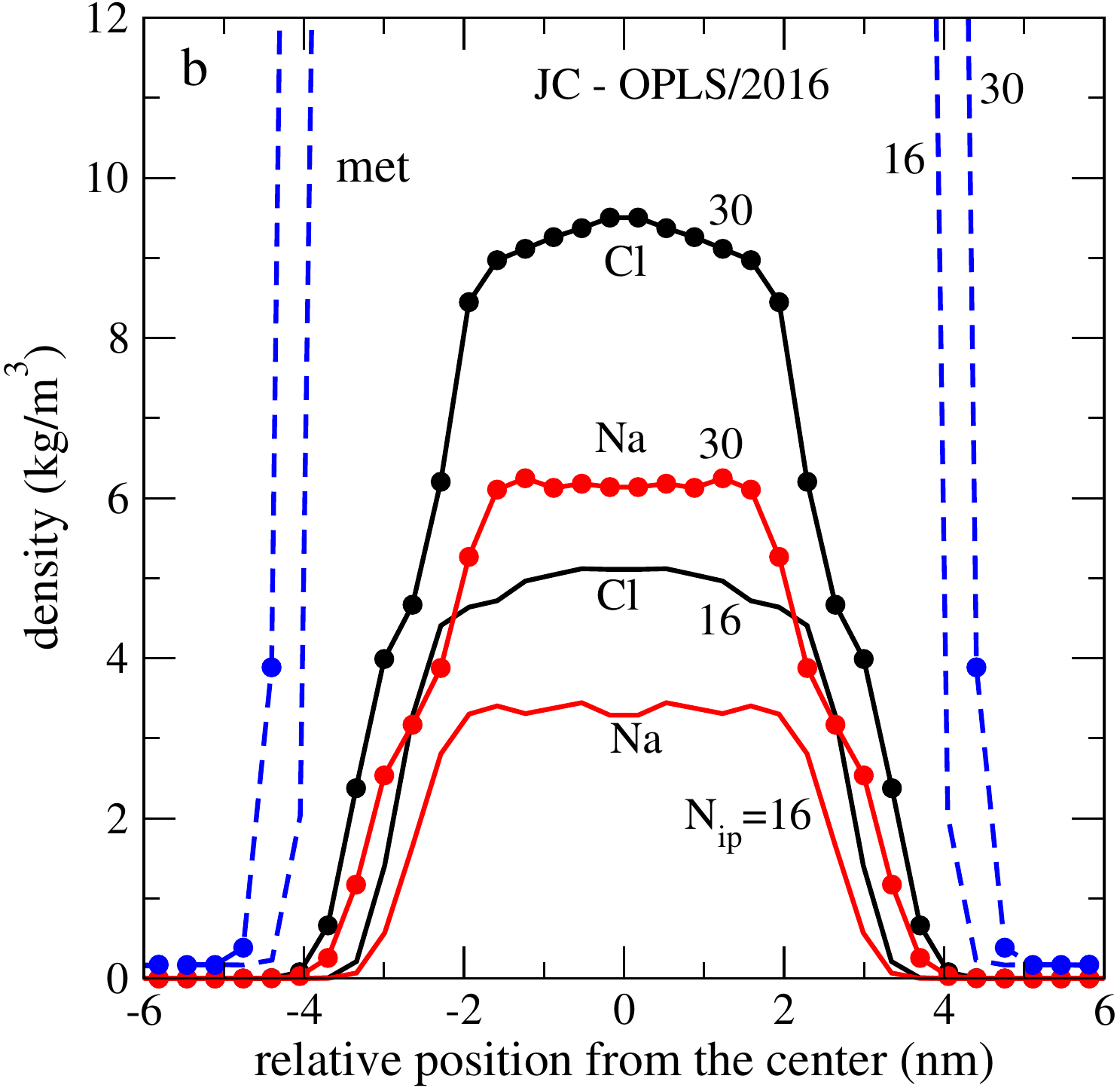}
\caption{(Colour online) 
Panel (a): Density profile of methanol molecules across the vapour-liquid interface for pure methanol and
for NaCl methanol solution at molality $m=0.3121$ within the JC-OPLS/2016 model.
Panel~(b): The same as in panel (a), but with the density profiles of ions, the number of NaCl ion pairs, $N_\text{ip}$,
is indicated in the figure.
}
\label{fig11}
\end{figure}

We performed a set of runs (of 8--10 pieces) at a constant volume, each piece with the time
 duration of 10~ns, and
obtained the result for $\gamma$ by taking the block average. However, only the results coming from
the blocks yielding most close $P_{xx}$ and $P_{yy}$ were taken into account. The results
are given in figure~\ref{fig10}. Our simulation data for pure methanol are shifted downward 
or say, systematically lower 
than the ones reported in \cite{vega}, presumably because of the difference 
in technical details of simulations, e.g., the geometric combination rule (CR3 in GROMACS
nomenclature) was used for OPLS/2016 model in~\cite{vega}. 
The experimental result for pure methanol
is at $\gamma=22.51$~mN/m.
However, the sequence of the magnitude of the results for different models
in the present work and in \cite{vega} is the same. The surface tension values
definitely increase with an increasing ion concentration. This trend is not strong, however.
From the density profiles of methanol species (the OPLS/2016 model is chosen
just for illustration) at $m=0$ and at $m=0.3121$, we conclude that the interface becomes
slightly sharper upon adding ions to methanol [figure~\ref{fig11}~(a)]. The density of the liquid slab is 
lower at both molalities than for the bulk system, cf. figure~\ref{fig1}~(a). Apparently, if the number
of particles in simulations increased, the interface width could slightly decrease and
the surface tension values might become higher.
On the other hand, as it follows from figure~\ref{fig11}~(b), the interface layer is constituted by
methanol molecules. There is no preferential adsorption of ions at the interface, as expected
for the nonpolarizable methanol and ions models from the macroscopic electrostatic arguments.  
The ions are expelled from the interface as witnessed by the density profiles of ions in figure~\ref{fig11}~(b).
This issue for aqueous solutions of salts was discussed in detail in~\cite{neyt}.

\section{Summary and conclusions}

We have performed a rather extensive set of molecular dynamic simulations 
to study various properties of NaCl solutions in methanol at room temperature
and ambient pressure. The models for solutions involve united atom
type description for methanol using L1, L2, TraPPE and OPLS/2016 force fields
and the JC, SD and VR models for ions. The combination of models of the solvent 
and solute species has been inspired by the previous developments for
aqueous solutions of NaCl.

We explored the evolution of density, surface tension, solvent dielectric constant and
self-diffusion coefficients of the species with an augmenting amount of NaCl solute
in the solution. The microscopic
structure is described in terms of the pair distribution functions,
the first coordination numbers of species and the number of the ion pairs.
Trends of behaviour of different properties of the system are determined by the balance
of effects resulting from non-electrostatic and electrostatic interactions. This is
manifested, for example, in a very different solubility of this salt in methanol solvent 
compared to water.

From a comparison with a limited set of experimental data for the systems in question and
with the results of other authors on related systems, we can conclude that the predictions
obtained are qualitatively correct and give a physically sound picture of the properties explored.
It seems that the OPLS/2016 model for methanol combined with JC force field for ions
performs best, compared to other combinations of the force fields.

At the present stage of the development, the missing and very important elements 
worth a much more detailed investigation are numerous.
1) Within the employed modelling, it is worth to explore the
shear viscosity and electric conductivity, as it has been done for
NaCl solutions in e.g.,~\cite{benavides1,benavides2,guevara,reiser}.
2) It is worth to explore the extension of the present models by involving either four-site or all-atom 
description of methanol combined with different force fields for ions.
3) Ampler insights into the trends of behaviour of various properties of this kind of solutions 
would follow from the study of a set of alkali halide salts in the same solvent.
4) The role of combination rules should be explored as well, see e.g.,~\cite{spohr} for
the case of aqueous solutions of NaCl.

The most challenging and remaining task, however, is to establish the solubility limit 
for the models in question by using the tools tested for NaCl aqueous solutions in 
\cite{benavides1,benavides2,aragones}. 
We hope to address some of these issues in a future work.

\section*{Acknowledgements}
M.C.S. and O.P. are grateful to M. Aguilar for technical support of this work
at the Institute of Chemistry of the UNAM. M.C.S. is grateful to CONACyT of Mexico
for Ph.D. scholarship.

\ukrainianpart

\title{Дослідження властивостей метанолового розчину NaCl методом молекулярної динаміки} 

\author{M. Круз Санчес\refaddr{label1}, Г. Домінгес\refaddr{label2}, O. Пізіо\refaddr{label3}}

\addresses{
	\addr{label1} Інститут хімії, Національний автономний університет м. Мехіко, Мехіко, Мексика
	\addr{label2} Інститут матеріалознавства, Національний автономний університет м. Мехіко, Мехіко, Мексика
	\addr{label3} Інститут хімії, Національний автономний університет м. Мехіко, Мехіко, Мексика
}

\makeukrtitle 

\begin{abstract}
	Симуляції методом молекулярної динаміка в ізотермічно-ізобаричному ансамблі застосовано для дослідження мікроскопічної структури та основних термодинамічних властивостей модельного розчину, щоскладається з солі  NaCl, розчиненої в метаноловому розчиннику.
	Задіяно чотири силові поля об'єднаного атома  для метанолу. Для іонних розчинів 
	використано моделі Джонга-Чітхема, Сміта-Данга, а також модель від лабораторії Врабека.   
	Основною метою є оцінка якості передбачень  різноманітних комбінацій моделей
	для базових властивостей цих розчинів. Зокрема, досліджено зміну густини від моляльності,
	структурні властивості в термінах різних парних функцій розподілу, координаційні числа, число іонних пар і середнє число водневих зв'язків.  Окрім цього, описано зміни коефіцієнтів самодифузії сортів, діелектричну сталу розчинника і еволюцію поверхневого натягу зі зміною концентрації іонів.

	\keywords  метанол, хлорид натрію, мікроскопічна структура, симуляції молекулярної динаміки
	
\end{abstract}

\end{document}